\documentclass[12pt]{article}

\usepackage{amsmath,amssymb,amsfonts}
\usepackage{latexsym}
\usepackage{bm}
\usepackage{bbm}
\usepackage{cite}
\usepackage{color}
\usepackage{graphicx}
\usepackage{slashed}
\usepackage{tabularx}
\usepackage{comment}

\newcommand{\Fig}[1]{Fig.~\ref{#1}}

\newcommand{\Eq}[1]{Eq.~(\ref{#1})}

\newcommand{\GeV}{\mbox{ GeV}}

\begin{document}

\title{\bf  The 130 GeV gamma-ray line and Sommerfeld enhancements}

\vspace{0.3truecm}
\author{
 Jing  Chen\footnote{Email: jchen@itp.ac.cn}
  \ and  Yu-Feng Zhou\footnote{Email: yfzhou@itp.ac.cn}
  \\ \\
 \textit{State Key Laboratory of Theoretical Physics},\\
  \textit{Kavli Institute for Theoretical Physics China,}\\
  \textit{Institute of  Theoretical Physics, Chinese Academy of Sciences,}\\
  \textit{Beijing, 100190, China}
}
\date{}
\maketitle
\begin{abstract}
  Recently, possible indications of line spectral features in the Fermi-LAT
  photon spectrum towards the galactic center have been reported.  If the
  distinct line features arise from dark matter (DM) annihilation into $\gamma
  X \ (X=\gamma, Z^{0} \text{ or } h^{0})$, the corresponding annihilation
  cross-section is unnaturally large for typical loop-induced radiative
  processes. On the other hand, it is still too small to be responsible for
  the observed DM relic density.  We show that the mechanism of Sommerfeld
  enhancement with scalar force-carrier can provide a simple solution to these
  puzzles. The possibly large Sommerfeld enhancement of the cross-section for $s$-wave DM
  annihilation can significantly reduce the required effective couplings
  between DM and charged particles in typical loop diagrams.  The DM particles
  necessarily annihilate into scalar force-carriers through tree-level
  $p$-wave process, which can dominate the total DM annihilation cross-section
  at freeze out, resulting in the correct thermal relic density, but has
  subdominant contributions to the DM annihilation today due to velocity suppression. We perform detailed
  analysis on the effects of $p$-wave Sommerfeld enhancement on freeze
  out. The results show that with the constraints from the thermal relic
  density, the required effective couplings can be reduced by an order of
  magnitude.
\end{abstract}

\newpage

\section{Introduction}
It has been well-established from observations that dark matter (DM)
contributes to nearly 23\% of the energy budget of the Universe. The leading
DM candidates such as the weakly interacting massive particles (WIMPs) can
interact weakly with the ordinary matter and possibly be detected through
direct and indirect searches.  Monoenergetic gamma-ray lines are one of the
smoking gun signals of halo DM annihilation, which is hard to mimic by
astrophysical sources.

Recently, a possible line spectral feature around 130 GeV or with an
additional line at 111 GeV in the Fermi-Large-Area-Telescope (Fermi-LAT)
photon spectrum in regions close to the galactic center (GC) have been
reported
\cite{1203.1312:Bringmann,1204.2797:Weniger,1205.1045:Tempel,1205.4723,1206.1616:Finkbeiner}.
Indications of similar spectral features have been reported in the galactic
clusters with modest statistical significance \cite{1207.4466:Raidal} and
possibly in unassociated gamma-ray point sources
\cite{1207.7060:Finkbeiner,1208.0828:Hooper,1208.1996:Tempel,1208.1693:Mirabal}.
The latest analysis from Fermi-LAT collaboration also shows the indication of such line feature
at 130 GeV (135 GeV) with 4.01 (3.35) $\sigma$ local significance using unreprocessed (reprocessed)
data \cite{FermiTalk130}. However, no globally significant gamma-ray line features  have been established from the analysis of Fermi-LAT collaboration.
It remains to be confirmed whether these line features are indeed from DM
annihilation or due to instrumental uncertainties \cite{1205.4700:Boyarsky}
or  astrophysical backgrounds \cite{1207.0458:Aharonian,1209.4548:Raidal,1209.4562:Finkbeiner}.
The upcoming HESS-II experiment can provide an independent check on
existence  of the gamma-ray line features.

The monoenergetic gamma-ray lines naturally arise from DM annihilation into
two-body final states $\gamma X$, where $X$ stands for the Standard Model (SM)
neutral particles $\gamma$, $Z^{0}$ and Higgs boson $h^{0}$ etc..  In this case, the energy of the photon is
given by $E_{\gamma}=m_{\chi}[1-m_{X}^{2}/(4m_{\chi}^{2})]$,  where $m_{\chi}$
is the DM mass. The observed line signals correspond to a thermally averaged cross
section multiplied by relative velocity
$\langle\sigma_{\gamma\gamma} v_{\text{rel}}\rangle\sim 1.27\times10^{-27}\text{ cm}^{3}\text{s}^{-1}$ for Einasto profile and
$\sim 2.27\times10^{-27}\text{ cm}^{3}\text{s}^{-1}$ for a generalized Navarro-Frenk-White (NFW) profile
respectively \cite{1204.2797:Weniger}.
Although the line spectral feature is expected from DM two-body annihilation,
the corresponding cross-section and non-observation of an excess of accompanying continuum gamma-ray flux may challenge the explanation in
terms of  simple WIMP models \cite{1205.6811:Hooper,1208.0009:Zurek,1208.4100:Bai}.
First, in most DM models the DM-photon couplings are generated radiatively
through loop level diagrams with charged intermediate states $F\bar F$ where
$F$ can be SM charged gauge bosons $W^{\pm}$ and charged fermions $f$ except
for the top-quark. For an annihilation cross-section
$\langle\sigma_{\gamma\gamma}
v_{\text{rel}}\rangle\approx\mathcal{O}(10^{-27})\text{ cm}^{3}\text{s}^{-1}$,
the required DM couplings to mediator particles are in general unnaturally
large, which may raise the question of perturbativity \cite{1205.6811:Hooper}.
Second, if the intermediate charged states are kinematically allowed to be the
annihilation final states such as $f\bar f$ and $W^{\pm}W^{\mp}$, the
corresponding tree level cross-sections are related to that of DM annihilation
into $2\gamma$ as $\langle\sigma_{f\bar{f},WW}
v_{\text{rel}}\rangle/\langle\sigma_{\gamma\gamma}
v_{\text{rel}}\rangle\sim(\pi/\alpha_{\text{em}})^{2}\approx2\times 10^{5}$.
Such a large cross-section is stringently  constrained by the non-observation of
any excesses in the continuum gamma-ray spectrum and the cosmic-ray antiproton
flux
\cite{1206.7056:Buchmuller,1207.0800:Slatyer,1207.1468:Cholis,1208.0267:Bi,
  1211.6739:Asano}.
Finally, if the $\chi\bar \chi \to F\bar F$ channels  are not opened and $\chi\bar{\chi}\to\gamma
X$ is the main DM annihilation
channel as suggested by the current observations, the corresponding cross-section is not
large enough to generate the correct DM thermal relic density which typically
requires $\langle\sigma v_{\text{rel}}\rangle_{F}\approx3\times10^{-26}\text{ cm}^{3}\text{s}^{-1}$ for $s$-wave
annihilation.

Several mechanisms have been proposed to address this problem such as
co-annihilation, forbidden channel, asymmetric DM \cite{1208.0009:Zurek},
resonant annihilation and cascade annihilation \cite{1208.4100:Bai},
etc.. Note that many of them introduce degeneracies in the mass of the DM
particles and the intermediate particles. Specific models in which the signals
of $\chi\bar{\chi}\to\gamma X$ and the thermal relic density are not
correlated have been considered in
Refs. \cite{1205.1520:x,1205.2688:x,1205.4675:x,1205.4151,1206.2279:x,1206.2863:x,1207.1341:x,1207.4981,1209.1093:x,1211.2835}.

In this work, we show that the Sommerfeld enhancement with \textit{scalar}
force-carrier can simultaneously explain the large loop level cross-sections
and the correct thermal relic density without introducing degeneracies in the
mass of the DM and mediator particles.  In the presence of Sommerfeld
enhancement, the cross-section for $s$-wave annihilation $\chi\bar{\chi}\to
\gamma X$ today can be greatly enhanced due to the multiple exchange of a
light force-carrier $\phi$, which reduces the required couplings in the loop
diagrams.  As the force-carriers are light, the DM particles necessarily
annihilate into the force-carriers.  For $\phi$ being a scalar particle, the
annihilation $\chi \bar \chi\to \phi\phi$ proceeds through $p$-wave. The
$p$-wave process is also Sommerfeld enhanced, which can dominate the total
annihilation cross-section at freeze out, lead to the correct thermal relic
density, but plays a subdominant role in the halo DM annihilation today due to
the velocity suppression. The total annihilation in the halo can be dominated
by $\gamma X$ channel, which can explain both the reported gamma-ray line
spectrum and the nonobservation of continuum spectrum.

This paper is organized as follows. In Sec.~\ref{sec:smf}, we begin with a
brief review on the main features of the Sommerfeld enhancement, and then focus
on the Sommerfeld enhancement of $p$-wave processes and its impact on the
thermal relic density. The effects of kinetic decoupling for $p$-wave processes
are also discussed.  In Sec.~\ref{sec:alpha-bound}, we discuss the maximally
allowed Sommerfeld enhancement factor after considering the constraints from
the thermal relic density for the case with scalar force-carriers. In Sec.
~\ref{sec:model}, we apply the mechanism to a reference DM model in which the
$\chi\bar\chi\to 2\gamma$ proceeds through one-loop diagrams and discuss
reduction of the required effective couplings between DM particles and the
charged intermediate states in the loop.  The conclusions are given in Sec.
\ref{sec:conclusion}.

\section{Sommerfeld enhancements of $p$-wave DM annihilation
  and thermal  relic density}\label{sec:smf}

\begin{figure}
  \begin{center}
    \includegraphics[width=0.32\textwidth]{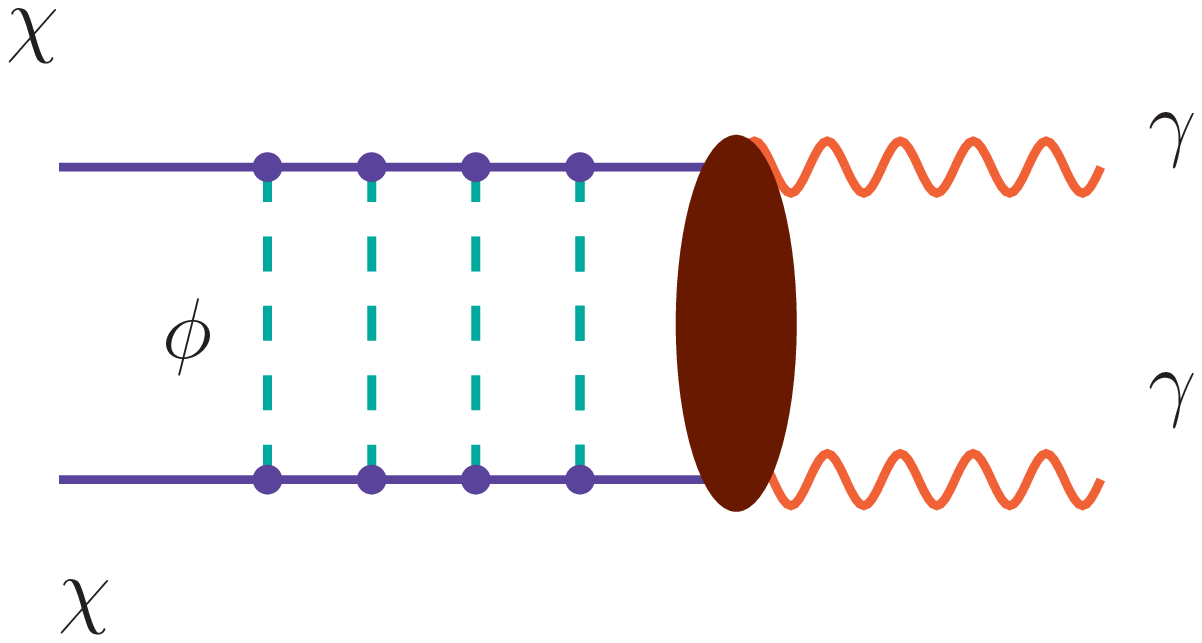}
      \includegraphics[width=0.32\textwidth]{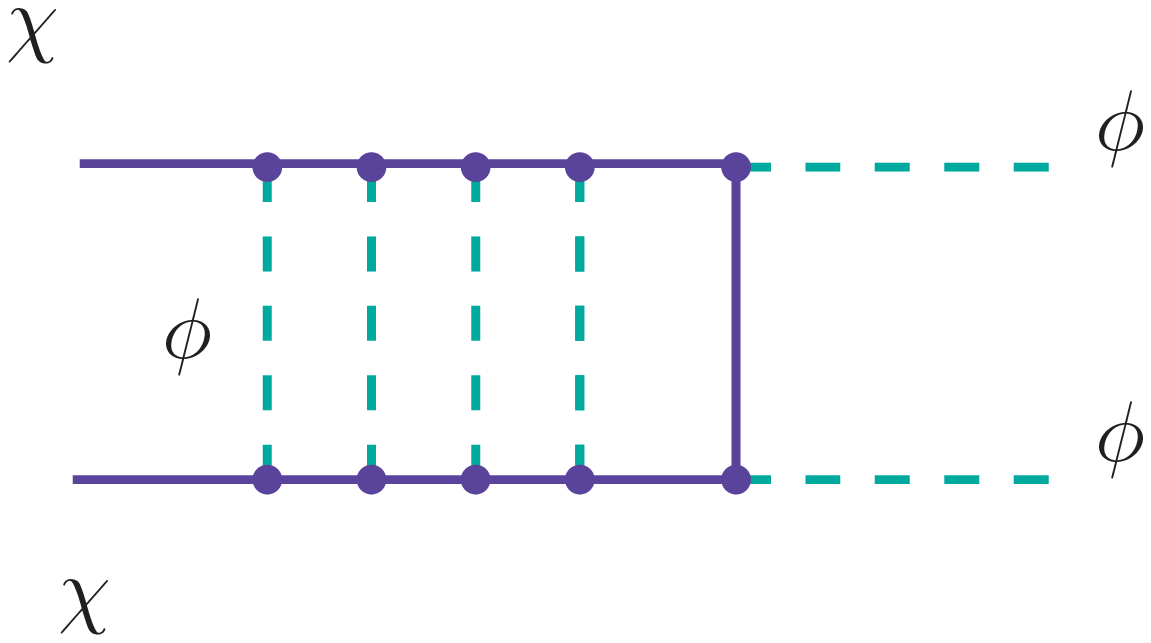}
    \includegraphics[width=0.32\textwidth]{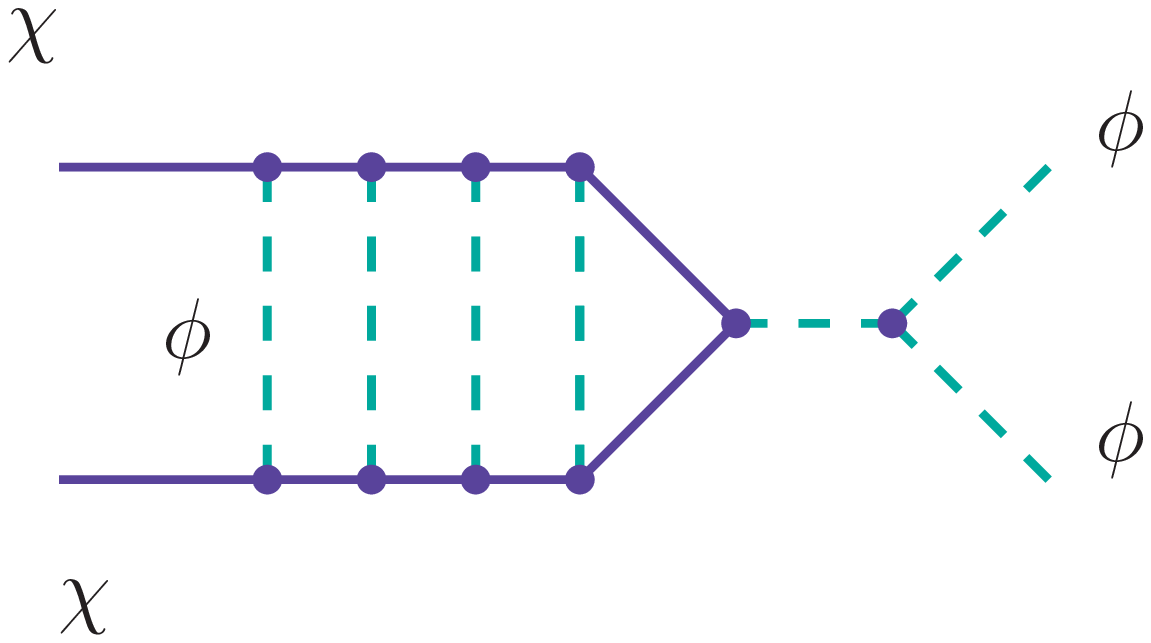}
    \caption{(Left) Feynman diagram of DM annihilation into $2\gamma$ with
      multiple-exchange of force-carriers which leads to the Sommerfeld
      enhancement; (Middle) diagram of $t$-channel DM annihilation into
      force-carrier particles; (Right) diagram of $s$-channel DM particle
      annihilation into force-carrier particles through $\phi^3$ type
      interactions.}
\label{fig:digrams}
\end{center}
\end{figure}

The Sommerfeld enhancement of annihilation cross-section occurs when the
annihilating particles self-interact through a long-range attractive potential
$V(\mathbf{r})$ at low velocities \cite{Sommerfeld31}. In this scenario, the
short-distance DM annihilation cross-section can be greatly enhanced due to
the distortion of the wave functions of annihilating particles at origin
\cite{Hisano:2002fk,Hisano:2003ec,Hisano:2004ds,Cirelli:2007xd}. The
attractive potential may originate from multiple-exchange of light force
carrier particles between the annihilating DM particles as shown in
Fig.~\ref{fig:digrams}. The nature of Sommerfeld enhancement have been
extensively studied (see, e.g.,
Refs.~\cite{ArkaniHamed:2008qn,Iengo:2009xf,Iengo:2009ni,Cassel:2009wt,Slatyer:2009vg,Hryczuk:2010zi,0909.4128:Dent,0910.5221:White,Feng:2009hw,Feng:2010zp})
in light of the cosmic-ray positron/electron excesses observed by
PAMELA~\cite{Adriani:2008zr}, ATIC \cite{Chang:2008zzr}, and Fermi-LAT
\cite{Abdo:2009zk} etc..

In the previous studies, the force-carrier $\phi$ is often assumed to be a
vector boson, as it can be naturally light~\cite{ArkaniHamed:2008qn}.  The
maximally allowed enhancement factor turned out to be stringently constrained
by the thermal relic density due to the additional $s$-wave annihilation
$\chi\chi\to \phi\phi$ in the presence of the light force-carrier $\phi$ ,
which may challenge the Sommerfeld enhancement as an explanation for those
excesses~\cite{Feng:2009hw,Feng:2010zp,1011.3082_Finkbeiner}.  Note that the
cosmic-ray lepton excesses may have astrophysical origins and may not be related to
the halo DM annihilation. In this work, we shall consider
the Sommerfeld enhancement as a mechanism to
simultaneously account for both the possible gamma-ray line signals and the
correct thermal relic density.  For this purpose, we shall focus on the case
of {\it scalar} force-carriers.  Although in both the vector and scalar cases
the induced long-range attractive potential is of the same Yukawa type.  For a
scalar force-carrier, the $\chi\bar\chi\to \phi\phi$ is dominated by $p$-wave
processes, which significantly modifies the relic density constraints due to
the different velocity dependencies.
The two-body wave function $\Psi(\mathbf{r})$ of the annihilating DM particles
satisfies the non-relativistic Schr$\ddot{\mbox{o}}$dinger equation
\begin{equation}
  -\frac{1}{m_{\chi}}\nabla^{2}\Psi(\mathbf{r})+V\left(\mathbf{r}\right)\Psi\left(\mathbf{r}\right)=m_{\chi}v^{2}\Psi\left(\mathbf{r}\right),
\end{equation}
where $v=v_{\text{rel}}/2$ is the velocity of DM particle in the
center-of-mass frame and $v_{\text{rel}}$ is the relative velocity of the
annihilating  DM particles.  After an expansion over angular momentum $\ell$, namely,
$\Psi(r,\theta)=\sum_{\ell}P_{\ell}(\cos\theta)\chi_{\ell}(r)/r$, the
Schr$\ddot{\mbox{o}}$dinger equation for radial wave function $\chi_{\ell}$
can be written as:
\begin{equation}\label{eq:shrodinger}
\frac{d^{2}\chi_{\ell}\left(t\right)}{dt^{2}}-\left[\frac{\ell\left(\ell+1\right)}{t^{2}}+\frac{V(t)}{m_{\chi} v^{2}}\right]\chi_{\ell}\left(t\right)+\chi_{\ell}\left(t\right)=0,
\end{equation}
where $t\equiv m_{\chi}vr$.
The above Schr$\ddot{\mbox{o}}$dinger equation can be
solved with the following boundary conditions \cite{Iengo:2009ni,Cassel:2009wt}
\begin{eqnarray}\label{eq:boundary}
\lim_{t\rightarrow0}\chi_{\ell}\left(t\right)  =  t^{\ell+1} \ \text{and}\
\lim_{t\rightarrow\infty}\chi_{\ell}\left(t\right)  \rightarrow C\sin\left(t-\frac{\ell\pi}{2}+\delta_{\ell}\right),
\end{eqnarray}
where $\delta_{\ell}$ is the phase shift and $C$ is a normalization
constant. With the above boundary conditions, the Sommerfeld enhancement factor
$S_{\ell}$ of the annihilation cross-section is given by
\cite{ArkaniHamed:2008qn}
\begin{equation}
  S_{\ell}\equiv\lim_{t\to0}\left|\frac{\chi_{\ell}\left(t\right)}{\chi_{\ell}^{(0)}\left(t\right)}\right|^{2}
=\left[\frac{(2\ell+1)!!}{C} \right]^{2} ,
\end{equation}
where $\chi_{\ell}^{(0)}(t)$ is the wave function in the free-motion case
without a potential.
The exchange of massive vector or scalar particles $\phi$ between the DM
particles results in an attractive Yukawa potential
\begin{equation}
  V(r)=-\frac{\alpha e^{-m_{\phi}r}}{r},
\end{equation}
where $\alpha=g^{2}/(4\pi)$ is the coupling strength of $\chi\bar{\chi}\phi$
or $\bar{\chi}\gamma^{\mu}\chi\phi_{\mu}$ type of interactions.
The nature of the Sommerfeld enhancement depends on the two variables
\begin{equation}
\epsilon_{v}\equiv \frac{v}{\alpha} \ \mbox{and} \ \epsilon_{\phi}\equiv \frac{m_{\phi}}{\alpha m_{\chi}} .
\end{equation}
In the limit of $\epsilon_{\phi}\ll \epsilon_{v}^{2}$, the Yukawa potential
can be well approximated by a Coulomb-type potential. The corresponding
Schr$\ddot{\mbox{o}}$dinger equation can be solved analytically for
arbitrary angular momentum and the enhancement factors are~\cite{Cassel:2009wt}
\begin{eqnarray}
  S_{\ell}^{\text{Col}} =\left\{
    \begin{aligned}
  \left(\frac{\pi}{\epsilon_{v}}\right)\frac{1}{1-\exp\left(-\pi/\epsilon_{v}\right)}, & \ \ \ (\text{for }\ell=0),
  \\
  S_{0}^{\text{Col}}\cdot\prod_{n=1}^{l}\left(1+\frac{1}{4n^{2}\epsilon_{v}^{2}}\right), &
  \ \ \ (\text{for }\ell  \neq  0).
\end{aligned}
\right.
\end{eqnarray}
For small $\epsilon_{v}/\pi\ll1$, the enhancement factors can be approximated by
$S_{\ell}^{\text{Col}}\approx2\pi/((2\epsilon_{v})^{2\ell+1}(\ell!)^{2})$.
Therefore, at low velocities the $s$- and $p$-wave Sommerfeld enhancement factor scales as
$1/v$ and $(1/v^{3})$ respectively.

In the case where $\epsilon_{\phi}$ is non-negligible, the $1/v$ behavior of
 $s$-wave cross-section breaks down. Through approximating the Yukawa potential by the
Hulth$\acute{\mbox{e}}$n potential, the $s$-wave Sommerfeld enhancement factor
can be estimated as \cite{Cassel:2009wt,Slatyer:2009vg}
\begin{equation}
S_{0}\approx\left(\frac{\pi}{\epsilon_{v}}\right)\frac{\sinh\left(\frac{2\pi\epsilon_{v}}{\pi^{2}\epsilon_{\phi}/6}\right)}{\cosh\left(\frac{2\pi\epsilon_{v}}{\pi^{2}\epsilon_{\phi}/6}\right)-\cos\left(2\pi\sqrt{\frac{1}{\pi^{2}\epsilon_{\phi}/6}-\frac{\epsilon_{v}^{2}}{\left(\pi^{2}\epsilon_{\phi}/6\right)^{2}}}\right)} .
\end{equation}
For $\epsilon_{\phi}\gg \epsilon_{v}$, namely, the deBroglie wavelength of
incoming particles is longer than the range of the interaction, the $s$-wave
Sommerfeld enhancement saturate with $S_{0}\sim 12/\epsilon_{\phi}$.  But for
some particular values of $\epsilon_{\phi}\simeq 6/(\pi^{2}n^{2}),
(n=1,2,3,\dots)$ for which the DM can form zero-energy bound states, there
exists additional resonant enhancements which scale as $1/v^{2}$. The resonant
enhancement is eventually cut off by the finite width of the resonance
\cite{ArkaniHamed:2008qn}.

\begin{figure}[thb]
  \begin{center}
    \includegraphics[width=0.49\textwidth]{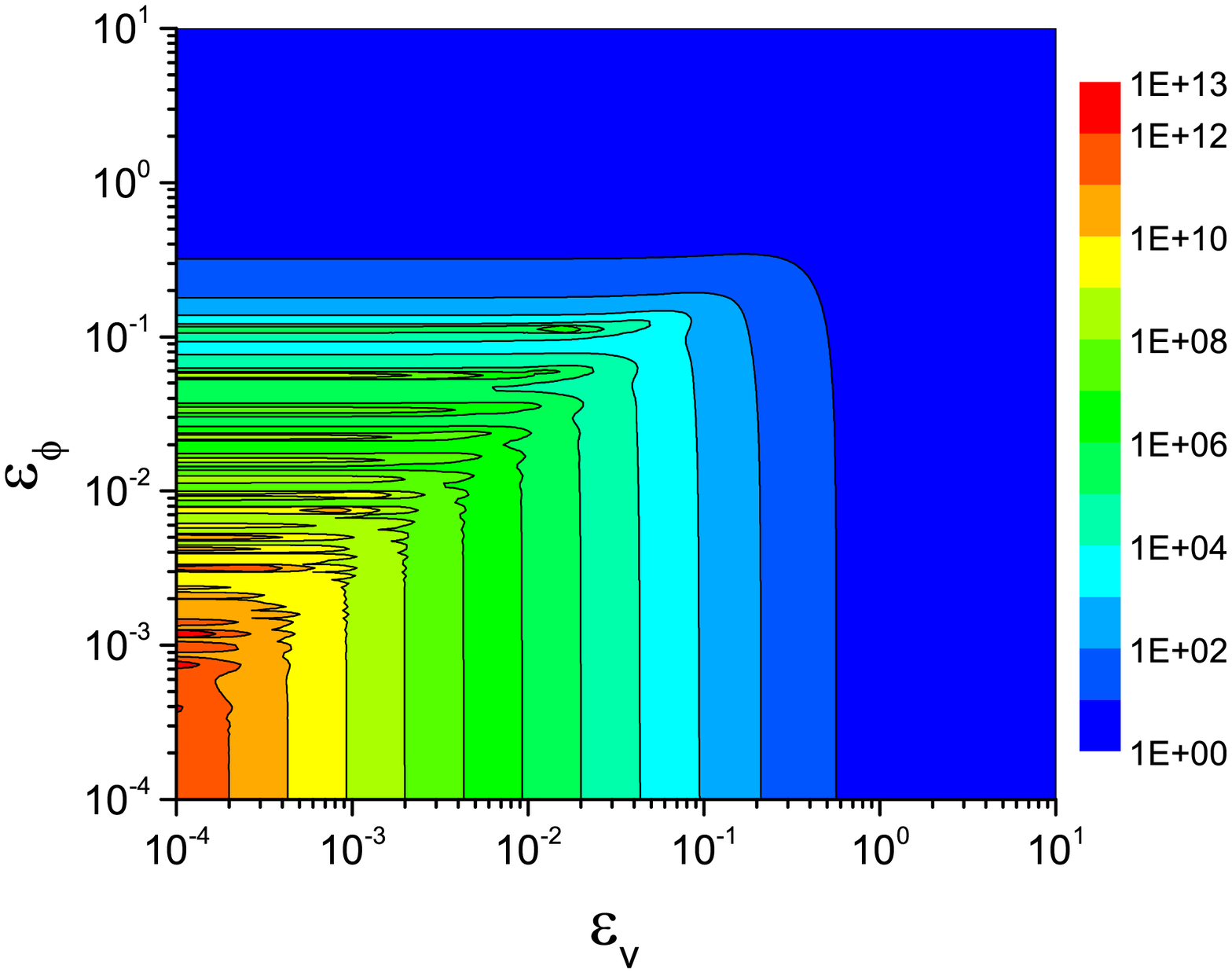}
    \includegraphics[width=0.49\textwidth]{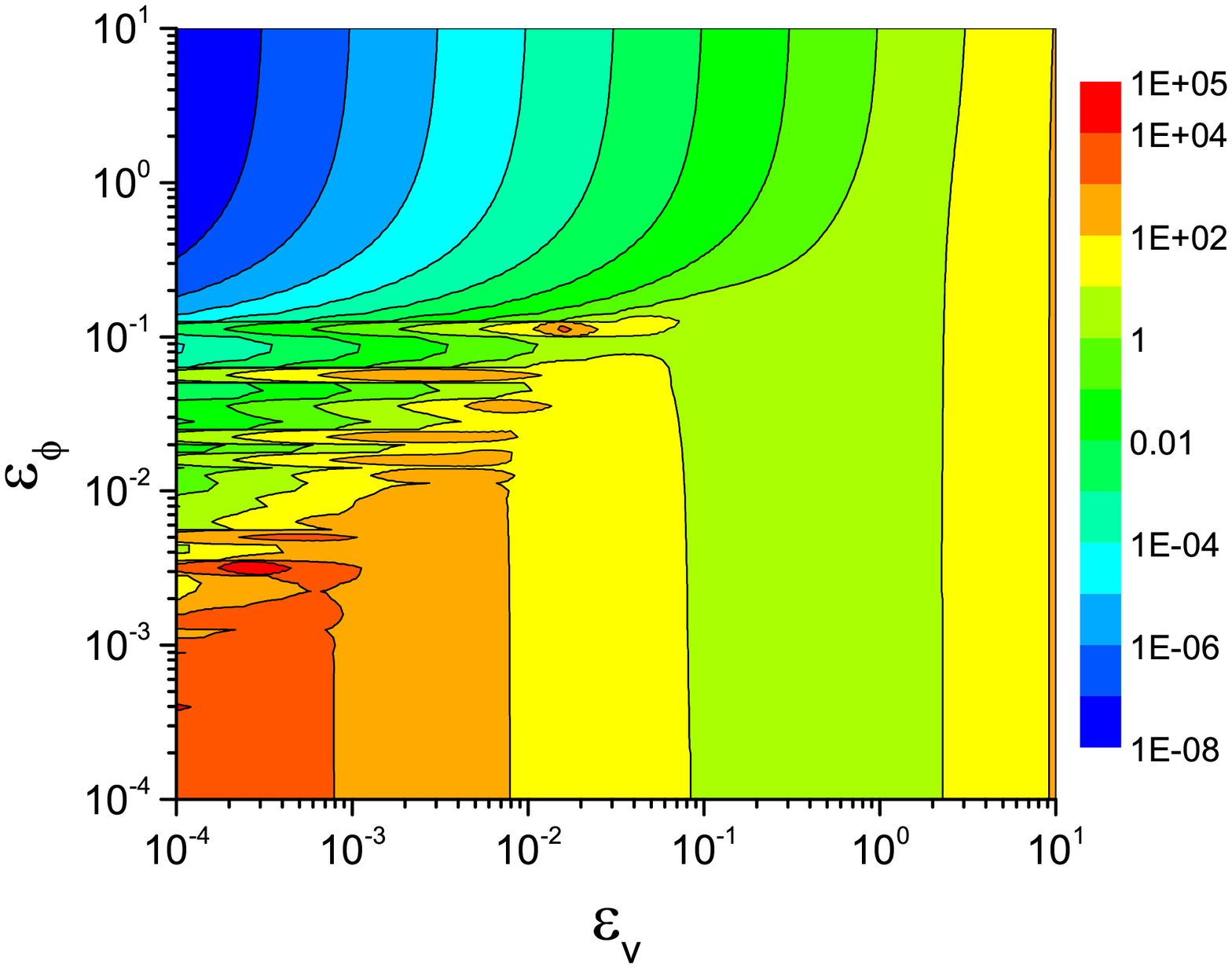}
    \caption{(Left) Contours of $p$-wave Sommerfeld enhancement $S_{1}$ with
      Yukawa potential in $(\epsilon_{v},\epsilon_{\phi})$ plane. (Right) the
      same but for $S_{1}\epsilon_{v}^{2}$ which is relevant to the total
      $p$-wave annihilation cross-section.}
\label{fig:contour-pp}
\end{center}
\end{figure}

For $p$-wave Sommerfeld enhancement factor $S_{1}$, there is no analytic
expression available. The $p$-wave case has been investigated in
Refs. \cite{Iengo:2009xf,Iengo:2009ni,Cassel:2009wt} without considering its
effects on the freeze out and thermal relic density.  In this work, we shall
focus on these effects as they are important in determining the maximally
allowed Sommerfeld enhancement.  We first numerically solve the
Eq. (\ref{eq:shrodinger}) with the boundary conditions in \Eq{eq:boundary} and
illustrate the dependence of $S_{1}$ on the two variables $\epsilon_{v}$ and
$\epsilon_{\phi}$ in \Fig{fig:contour-pp}.
For small $\epsilon_{\phi} \lesssim 10^{-3}$, the value of $S_{1}$ scales as
$1/v^{3}$ as expected from the Coulomb limit. For larger $\epsilon_{\phi}$ in
the range $10^{-3}-10^{-1}$, it has resonant behavior which is similar to the
$s$-wave case.  For even larger $\epsilon_{\phi} \gtrsim10^{-1}$ the
enhancement saturates.  Note that the generic $p$-wave annihilation
cross-section before including the Sommerfeld enhancement is proportional to
$v^{2}$, thus the velocity dependence of the total Sommerfeld enhanced
$p$-wave annihilation cross-section should be proportional to
$S_{1}\epsilon_{v}^{2}$.  In the right panel of Fig. \ref{fig:contour-pp}, the
contours of $S_{1}\epsilon_{v}^{2}$ in the $(\epsilon_{v},\epsilon_{\phi})$
plane are shown.  In the region where $\epsilon_{\phi} \lesssim 10^{-3}$, it
scales as $1/v$ instead of $1/v^{3}$.  In the resonance region
$10^{-3}\lesssim\epsilon_{\phi} \lesssim 10^{-1}$, the velocity dependence of
$S_{1}\epsilon_{v}^{2}$ is not significant. In the saturation region
$\epsilon_{\phi} \gtrsim 10^{-1}$, $S_{1}\epsilon_{v}^{2}$ scales as $v^{2}$,
the total cross-section decreases rapidly towards low velocities. Thus the
main difference from the $s$-wave case is that the total $p$-wave annihilation
cross-section can be either velocity-suppressed or velocity-enhanced, depending
on the size of $\epsilon_{\phi}$.

An important quantity directly related to the thermal relic density is
thermally averaged annihilation cross-section.  The generic DM annihilation
cross-section times the relative velocity before including the Sommerfeld
enhancement has the form $(\sigma
v_{\text{rel}})_{0}=a+bv_{\text{rel}}^{2}+\mathcal{O}(v_{\text{rel}}^{4})$,
where $a$ and $b$ are coefficients correspond to the $s$- and $p$-wave
contributions which are velocity-independent. After including the Sommerfeld
enhancement,
the thermally averaged cross-section at a temperature $T$ can be written as
\begin{equation}
\left\langle \sigma v_{\text{rel}}\right\rangle =a\langle S_{0}(v_{\text{rel}})\rangle+b\langle v_{\text{rel}}^{2}S_{1}(v_{\text{rel}})\rangle,
\end{equation}
where the thermal average of a quantity $\mathcal{X}(v_{\text{rel}})$ in the non-relativistic limit is given by
\begin{equation}
  \left\langle \mathcal{X}\right\rangle =\frac{x^{3/2}}{2\sqrt{\pi}}\int_{0}^{\infty} \mathcal{X}(v_{\text{rel}})
  e^{-\frac{xv_{\text{rel}}^{2}}{4}}v_{\text{rel}}^{2}\ dv_{\text{rel}},
\end{equation}
with $x \equiv m_{\chi}/T$. The thermally averaged annihilation cross-section
is a function of $x$ and depends on the parameters $\alpha$ and $m_{\phi}$.

The time evolution of the DM number density $n_{\chi}$ is governed
by the Boltzmann equation
\begin{equation}\label{eq:Boltzmann}
  \frac{dn_{\chi}}{dt}+3Hn_{\chi}=-\left\langle \sigma v_{\text{rel}}\right\rangle
  \left[n_{\chi}^{2}   -(n_{\chi}^{\text{eq}})^{2}\right] ,
\end{equation}
where  $n_{\chi}^{\text{eq}}$ is the equilibrium DM
number density and $H$ is the Hubble constant. The equation is often  rewritten as
\begin{equation} \label{eq:Boltzmann2}
\frac{dY}{dx}=-\sqrt{\frac{\pi}{45}}m_{\text{Pl}}m_{\chi}\frac{g_{*s} g_{*}^{-1/2}}{x^{2}}\left\langle \sigma v_{\text{rel}}\right\rangle \left[ Y^{2}-(Y^{\text{eq}})^{2}\right] ,
\end{equation}
where $Y^{(\text{eq})}\equiv n_{\chi}^{(\text{eq})}/s$ is the (equilibrium)
number density rescaled by entropy density $s$,
$m_{\text{Pl}}\simeq1.22\times10^{19}\text{ GeV}$ is the Planck mass scale.
$g_{*s}$ and $g_{*}$ are the effective relativistic degrees of freedom for
entropy and energy density respectively.
The decoupling temperature $x_{f}$ is defined as the temperature at which the dark matter
particles start to depart from the thermal equilibrium, and  the density $Y$ is
related to the equilibrium density $Y^{\text{eq}}$ by $Y\left(x_{f}\right) \equiv (1+c)Y^{\text{eq}}\left(x_{f}\right)$, where $c$ is a constant of order unity.
The value of $x_{f}$ is  approximately given by \cite{PhysRevD.33.1585}
\begin{eqnarray}
x_{f} &\approx &\ln[0.038c(c+2)m_{\text{Pl}}m_{\chi}(g_{\chi}g_{*}^{-1/2})\langle \sigma v_{\text{rel}}\rangle ]\nonumber\\
& & -\frac{1}{2}\ln\ln[0.038c(c+2)m_{\text{Pl}}m_{\chi}(g_{\chi}g_{*}^{-1/2})\langle \sigma v_{\text{rel}}\rangle ] ,
\end{eqnarray}
with $g_{\chi}$ the degrees of freedom of dark matter particle. In the
absence of Sommerfeld enhancement $\langle \sigma v_{\text{rel}}\rangle=a+6 b/x_{f}$, taking $c\approx1(2)$ for $s(p)$-wave annihilation
leads to  good fits to the numerical solutions of the Boltzmann equation.  The DM number density in the present-day
can be obtained by integrating  Eq.~(\ref{eq:Boltzmann2}) with respect to $x$ in the region
$x_{f}\leq x \leq x_{s}$,  where $x_{s}$ corresponds to the temperature at which the DM
annihilation rate is insignificant compared with that of the expansion of the Universe  and $Y(x)$ becomes stable.
The value of $Y(x_{s})$ can be written as
\begin{eqnarray}
  \frac{1}{Y\left(x_{s}\right)}
  & = & \frac{1}{Y\left(x_{f}\right)}+\sqrt{\frac{\pi}{45}}\ m_{\text{Pl}}\ m_{\chi}\int_{x_{f}}^{x_{s}}\frac{g_{*s}g_{*}^{-1/2}}{x^{2}}
  \left\langle \sigma v_{\text{rel}}\right\rangle dx .
\end{eqnarray}
In performing the integration over $x$, as $\left\langle \sigma
  v_{\text{rel}}\right\rangle$ depends on temperature, one needs to take into
account the effects of kinetic decoupling.  When the DM particles are in both
chemical and kinetic equilibrium with the radiation background, the
temperature of the DM particles tracks that of the background, i.e.,
$T_{\chi}=T$ or $x_{\chi}\equiv m_{\chi}/T_\chi=x$. After dropping out of
chemical equilibrium, the DM particles can still remain in kinetic equilibrium
with the radiation background through scattering off SM relativistic particles
which are in thermal equilibrium with the radiation background. At some
temperature $T_{\text{kd}}$, when the rate of the scattering cannot compete
with that of the expansion of the Universe, the DM particles start to
decouple from kinetic equilibrium.  After the kinetic decoupling, $T_{\chi}$
drops quickly with the scale factor $a$ as $a^{-2}$ instead of $a^{-1}$, the
temperatures of the DM particles and the radiation background are
approximately related by $T_{\chi}=T^{2}/T_{\text{kd}}$, or
$x_{\chi}=x^{2}/x_{\text{kd}}$~\cite{Bringmann:2006mu}.  Thus the integration
from $x_{f}$ to $x_{s}$ needs to be separated into two parts, from $x_f$ to
$x_{\text{kd}}$ and from $x_{\text{kd}}$ to $x_{s}$. For the second part of
the integration, one should use $x_{\chi}$ instead of $x$.  Previous analysis
have shown that including the effect of kinetic decoupling leads to a
significant reduction of relic density in the case of $s$-wave annihilation
with Sommerfeld
enhancement~\cite{0909.4128:Dent,0910.5221:White,Feng:2010zp}.
The value of $T_{\text{kd}}$ is model-dependent, for instance, in
supersymmetric models $T_{\text{kd}}\approx
10^{-3}-10^{-1}T_{f}$~\cite{0903.0189_Bringmann}. In this work, we shall take
the value of $T_{\text{kd}}$ as a free parameter. Finally, after freeze out,
the relic abundance of DM particles is given by
\begin{equation}
  \Omega h^{2}\approx2.76\times10^{8} Y \left(x_{s}\right)\left(\frac{m_{\chi}}{\text{GeV}}\right),
  \end{equation}
 which is to be compared with the observed value $\Omega h^{2}=0.113\pm0.004$~\cite{1001.4538_WMAP7}.

\begin{figure}[thb]
  \begin{center}
    \includegraphics[width=0.49\textwidth]{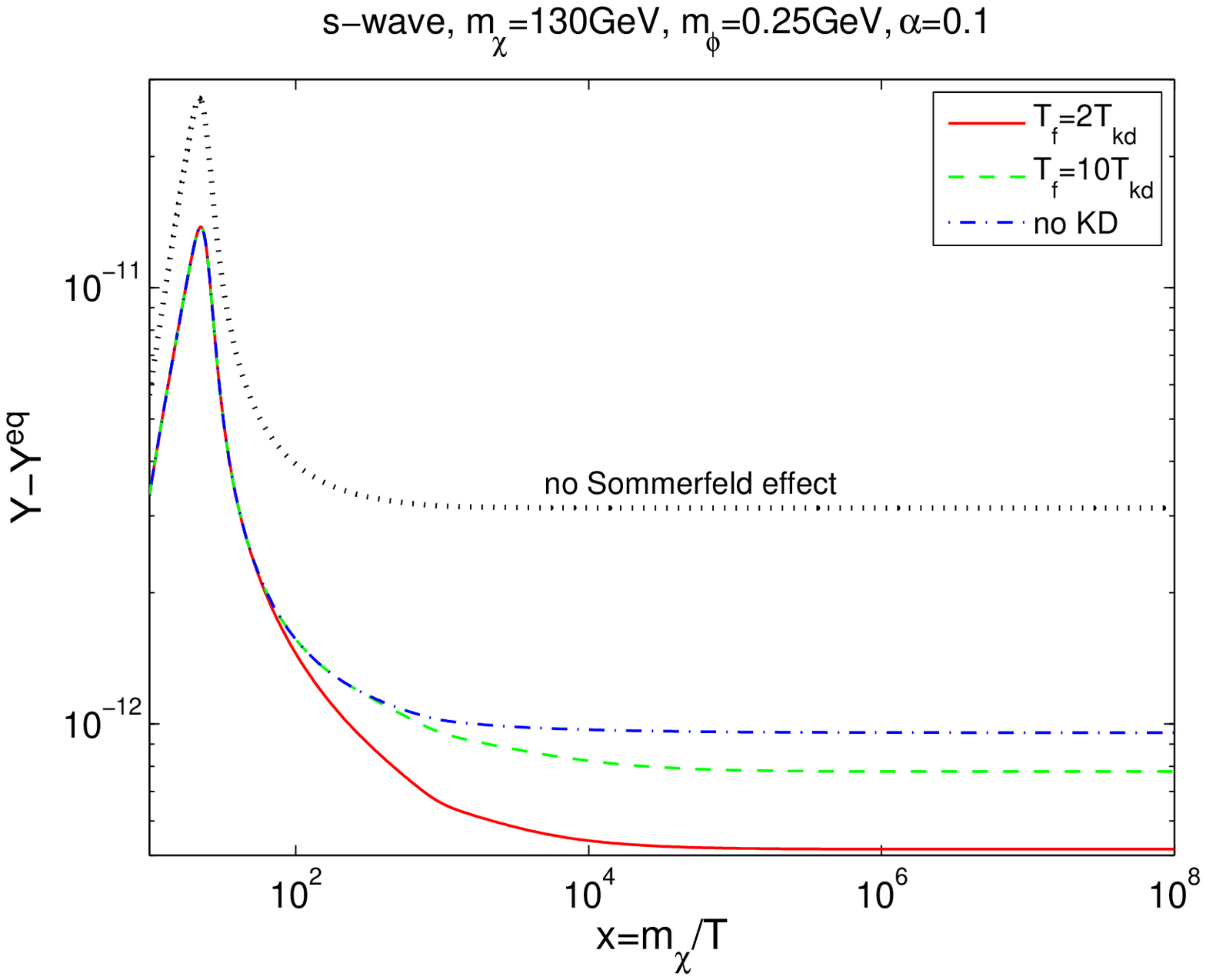}
    \includegraphics[width=0.49\textwidth]{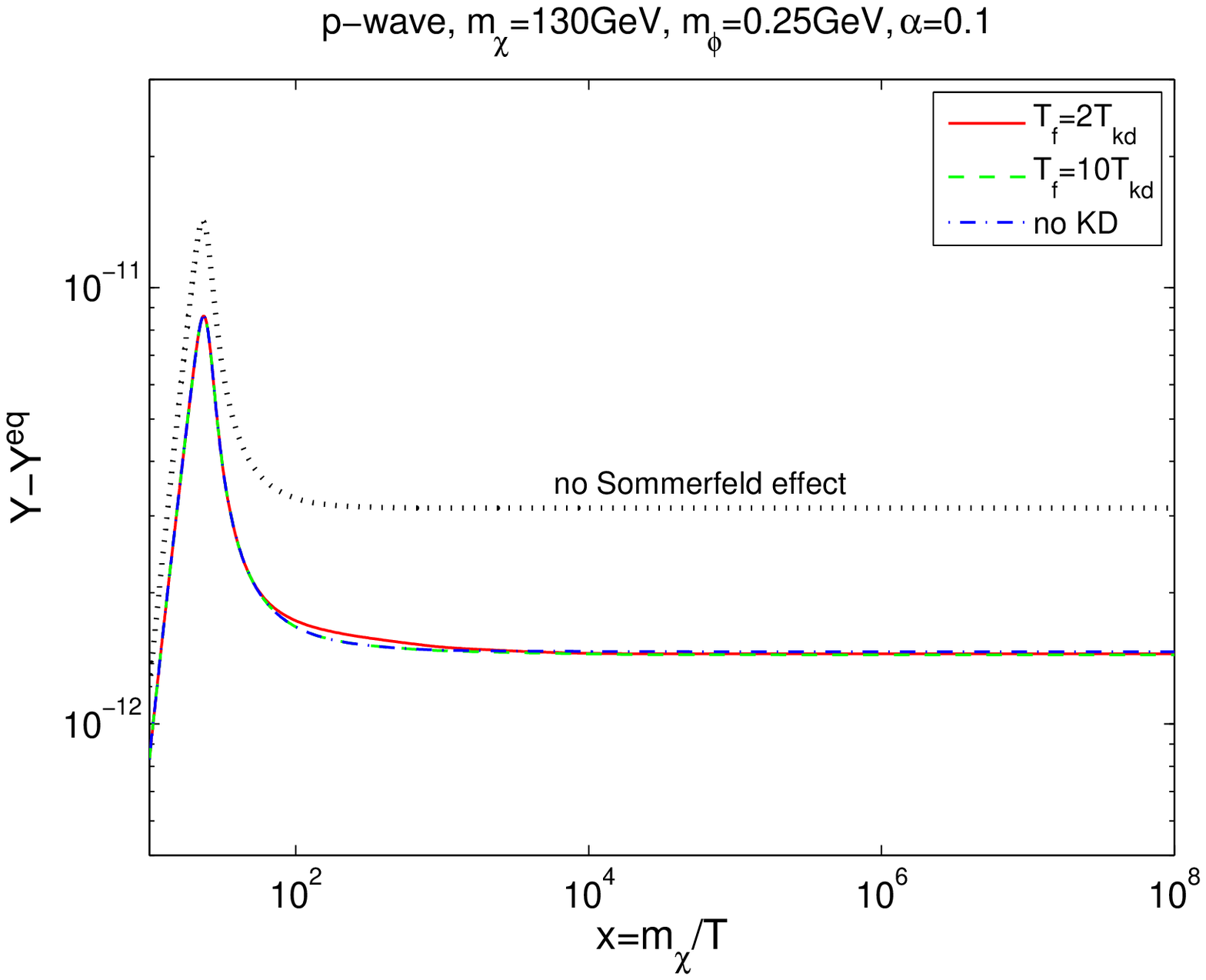}
    \includegraphics[width=0.49\textwidth]{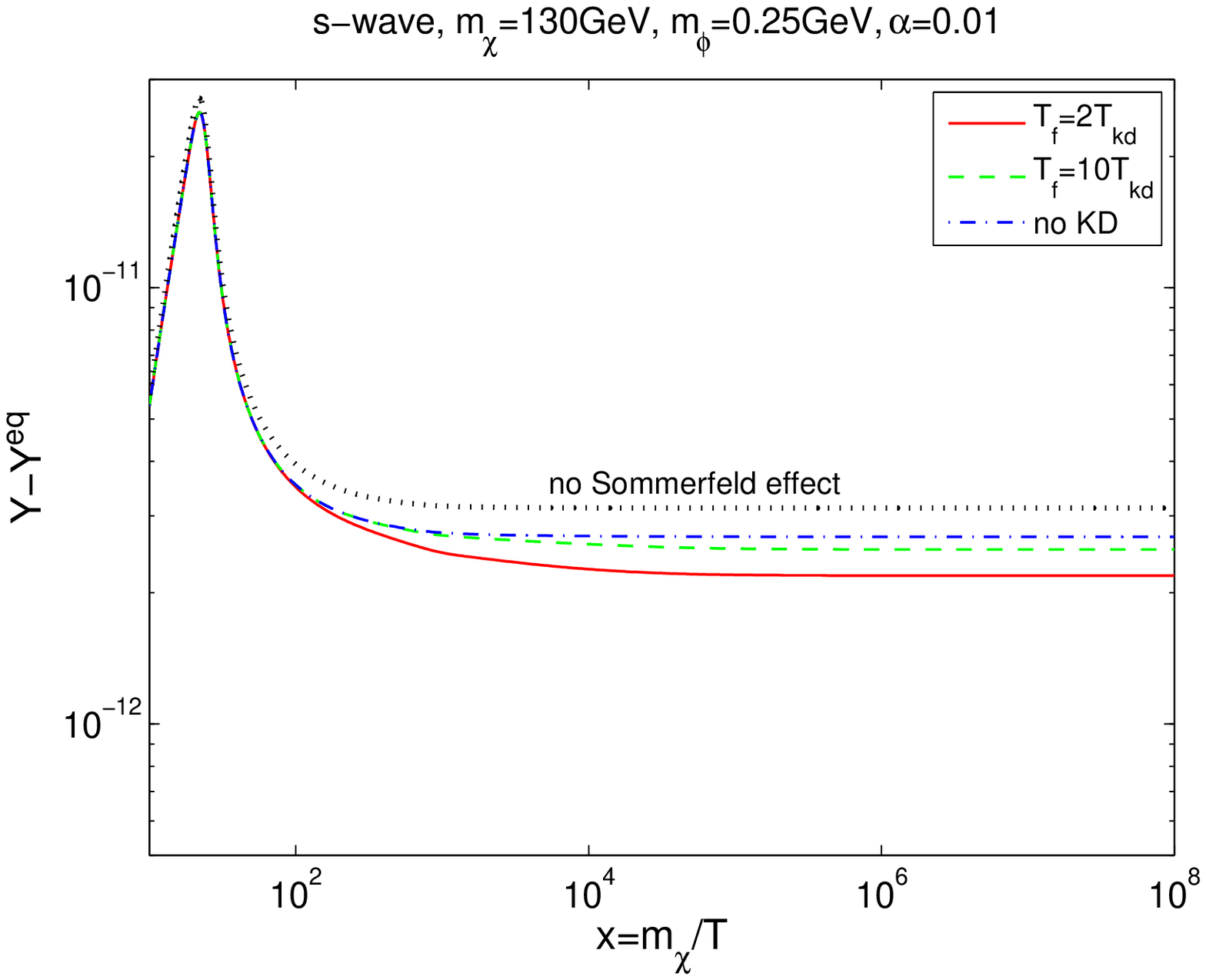}
    \includegraphics[width=0.49\textwidth]{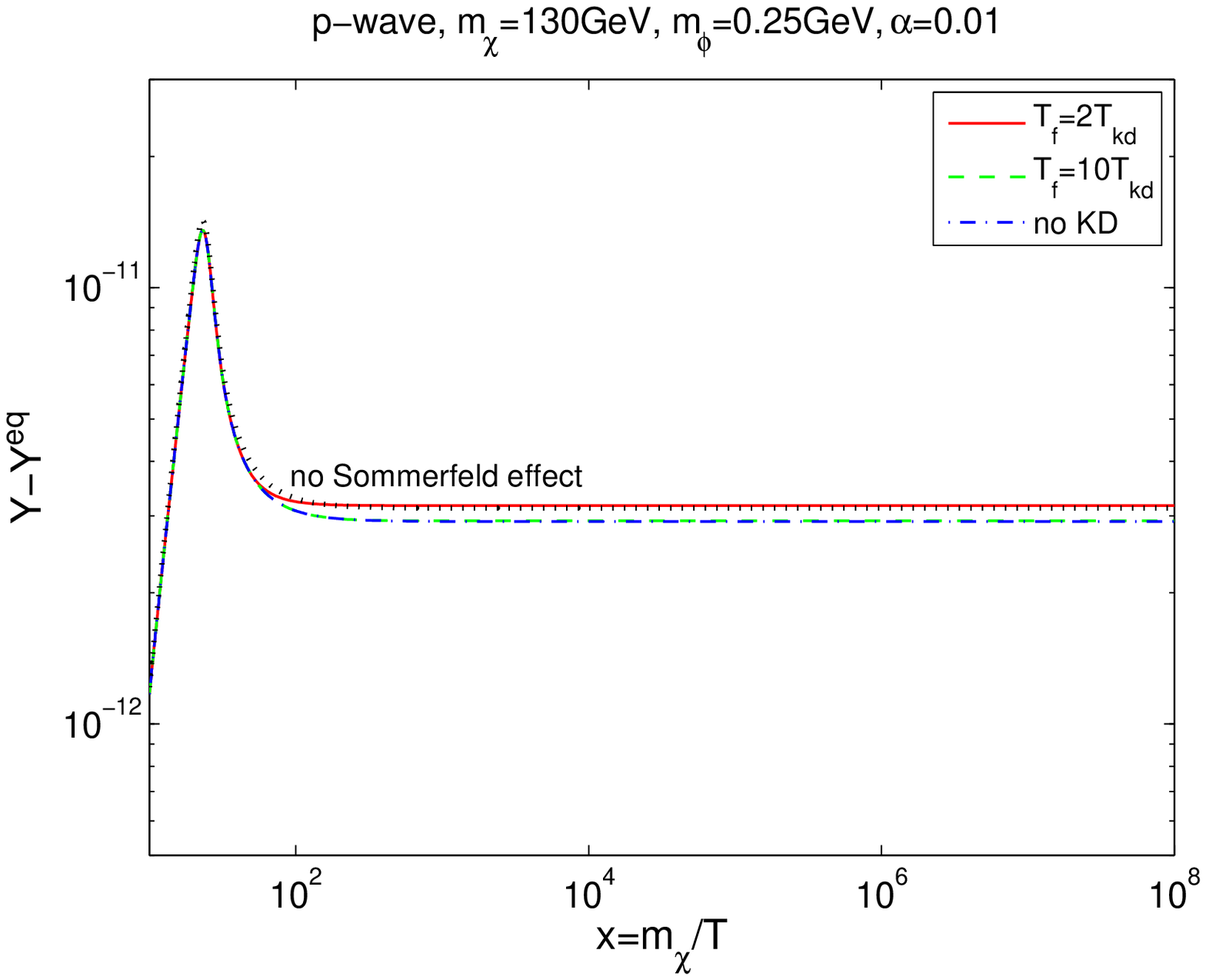}
    \caption{Effects of Sommerfeld enhancement and kinetic decoupling on the temperature evolution of the
      DM number density for the case of $s$-wave (left) and $p$-wave (right)
      with $\alpha=$0.1 and 0.01, see text for explanations. }
    \label{fig:kindec}
\end{center}
\end{figure}

 We solve the Boltzmann equation Eq.(\ref{eq:Boltzmann2})
 numerically including the effects of Sommerfeld enhancement and kinetic
 decoupling for both $s$-wave and $p$-wave annihilation.  In order to
 facilitate the comparison, we take $a=2.2\times10^{-26}\text{
   cm}^{3}\text{s}^{-1}$ and $b=1.7\times10^{-25}\text{
   cm}^{3}\text{s}^{-1}$ such that the final DM relic abundance $\Omega h^{2}\approx 0.113$ is the same
 for both $s$-wave and $p$-wave in the absence of Sommerfeld enhancement and
 kinetic decoupling.
 We fix $m_{\chi}=130\GeV$ and $m_{\phi}=0.25\GeV$ and consider two different
 values of $\alpha=0.1$ and $0.01$, and two kinetic decoupling temperatures
 $T_{f}=2T_{\text{kd}}$ and $10T_{\text{kd}}$.  The evolutions of
 $Y(x)-Y^{eq}(x)$ as a function of $x$ are shown in \Fig{fig:kindec}.  For
 both $s$- and $p$-wave annihilation, the inclusion of Sommerfeld enhancement
 results in reduction of final thermal relic density by a factor of
 $\mathcal{O}(1)$.  For $\alpha=0.1$ the relic density is reduced by a factor
 of $\sim 3$ for $s$-wave annihilation while $\sim 2$ for $p$-wave case.  At
 high temperatures $x\sim 20$, the $p$-wave annihilation cross-section still
 decreases with temperature in the presence of Sommerfeld enhancement, which
 leads to earlier decoupling from the thermal background and larger relic
 density.
 As it can be seen in \Fig{fig:kindec}, the effects of kinetic decoupling are
 significantly different between $s$- and $p$-wave case.  For
 $T_{f}=2 T_{\text{kd}}$ and $ 10 T_{\text{kd}}$, the kinetic decoupling leads to further
 reduction of relic density by $\sim 50\%(20\%)$ in the $s$-wave
 case. However, for $p$-wave case, the reduction is almost invisible for
 $\alpha=0.1$ as shown in Fig.~\ref{fig:kindec}. For smaller $\alpha=0.01$, it
 even leads to a slight enhancement of the relic density. This is because the
 kinetic decoupling makes the DM particle freeze out more quickly in the
 $p$-wave case as the annihilation cross-section decreases with temperature.

\section{Sommerfeld enhancement of $\chi\bar\chi\to \gamma X$ cross-sections }\label{sec:alpha-bound}
With the main features of $p$-wave Sommerfeld enhancement given in the previous section,
we are ready to discuss
the contribution to the DM thermal relic density from
the process of DM particles annihilating into scalar force-carriers through $p$-wave.
Since the force-carrier is  much lighter than the DM particle, i.e., $m_{\phi}\ll m_{\chi}$,
the DM particles necessarily annihilate into the force-carriers,
which contributes to a DM annihilation channel in addition to  $\gamma X$,
and can be the dominant contribution to
the total DM annihilation cross-section.
The process itself is also Sommerfeld-enhanced,
which complicates the calculations.

Before including the effects of Sommerfeld enhancements,
the total DM annihilation cross-section $ (\sigma_{\text{tot}}v_{\text{rel}})_0$ can be written as
\begin{equation}
  (\sigma_{\text{tot}} v_{\text{\text{rel}}})_{0}
  =(\sigma_{\phi\phi}v_{\text{rel}})_{0}+(\sigma_{\gamma X}v_{\text{rel}})_{0} .
\end{equation}
The DM particles can always pair-annihilate into $\phi\phi$ through
$t$-channel $\chi$-exchange.
In addition,
if there exists non-negligible cubic and quartic self-interactions
between the force-carriers of the form
$ -\mu \phi^{3}/3!-\lambda \phi^{4}/4!$,
two-body or three-body $s$-channel annihilation may occur.
The three-body annihilation is highly suppressed by small phase-space and
is neglected in this work.
In the case of  scalar force-carrier,
both the $t$- and $s$-channel two-body annihilation into $\phi\phi$
corresponding to the two diagrams in \Fig{fig:digrams}
are $p$-wave processes.   
In the limit of $m_\phi \ll m_\chi$,
the total annihilation cross-section of
$\chi\bar\chi\to \phi\phi$ from
the  calculation of the two diagrams in \Fig{fig:digrams} is independent of $m_\phi$,
and is given by
\begin{equation}\label{eq:phi-phi}
  (\sigma_{\phi\phi}v_{\text{rel}})_{0} =\frac{3\pi\alpha^{2}}{8m_{\chi}^{2}}
  \left(
  1-\frac{5}{18}\xi +\frac{1}{48}\xi^{2}
  \right)
  v_{\text{rel}}^{2},
\end{equation}
with $\xi=\mu/(2m_{\chi }\sqrt{\alpha \pi})$.
The first term in the right-hand-side of \Eq{eq:phi-phi} comes from the $t$-channel contribution. The term proportional to $\xi^2$ corresponds to the squared amplitude of $s$-channel annihilation,
and the term proportional to $\xi$ corresponds to the interference between $s$- and $t$-channel diagrams.
We assume that
the other annihilation channel  $\chi\bar\chi\to\gamma X$ is an $s$-wave process
such that there is no explicit velocity dependence in
$(\sigma_{\gamma X}v_{\text{rel}})_{0}$.
After including the Sommerfeld enhancement,
the total thermally averaged annihilation cross-section is
\begin{equation} \label{eq:crosssection}
  \langle \sigma_{\text{tot}} v_{\text{rel}} \rangle=
   (\sigma_{\gamma X} v_{\text{rel}})_{0} \langle S_{0}\rangle
  + \langle (\sigma_{\phi\phi} v_{\text{rel}})_{0} S_{1}\rangle.
\end{equation}
The Sommerfeld enhancement factors $S_{0,1}$ depend on parameters $\alpha$ and $m_\phi$,
and the cross-section $(\sigma_{\phi\phi} v_{\text{rel}})_{0} $  is  also a function of  $\alpha$.
Therefore,
the requirement of reproducing the correct thermal relic density
determines  the  size  of $\alpha$ for
a given $ (\sigma_{\gamma X} v_{\text{rel}})_{0}$
and  force-carrier mass $m_\phi$.
Using the value of  $\alpha$ constrained by the thermal relic density,
one can calculate  the allowed Sommerfeld enhancement factor $\langle S_0 \rangle$
for  $(\sigma_{\gamma X} v_{\text{rel}})_{0}$ at low temperatures,
which is needed to predict the cross-section of  $ \chi\bar\chi\to \gamma X$
in the present day.
\begin{figure}[thb]
  \begin{center}
    \includegraphics[width=0.49\textwidth]{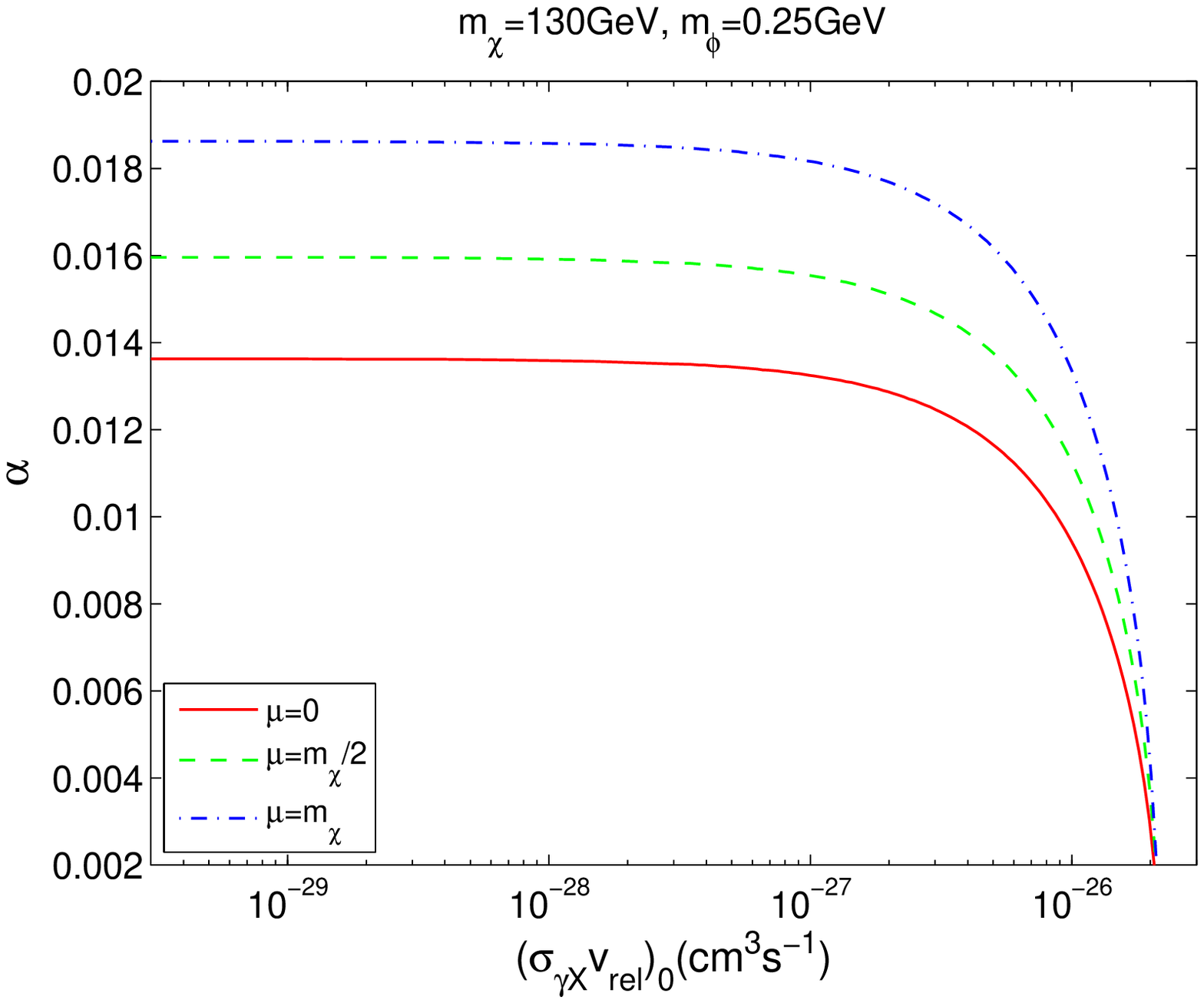}
     \includegraphics[width=0.49\textwidth]{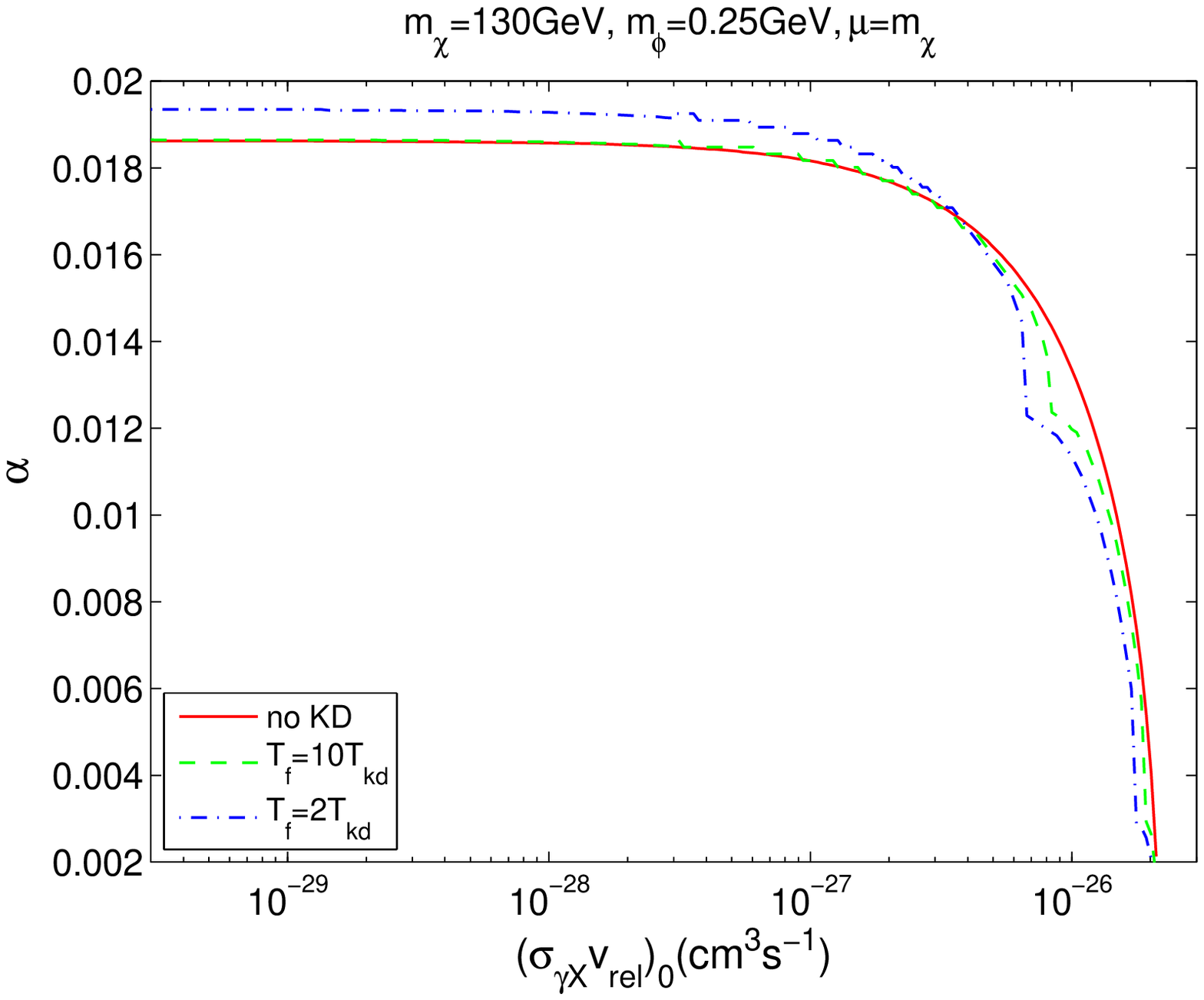}
     \caption{ The value of $\alpha$ allowed by the DM thermal relic density
       as a function of $(\sigma_{\gamma X}v_{\text{rel}})_{0}$. (Left)
       the cases without kinetic decoupling.  Three curves correspond to
       $\mu=0$ (solid), $m_\chi/2$ (dashed), and $m_\chi$ (dot-dashed),
       respectively.  (Right) the case with kinetic decoupling. Three curves correspond to without kinetic decoupling
       (solid), $T_{\text{kd}}=T_f/10$ (dashed) and $T_{f}/2$ (dot-dashed),
       respectively with  $\mu$ fixed at $m_{\chi}$. }\label{fig:constraints}
\end{center}
\end{figure}
%

In the left panel of \Fig{fig:constraints},
we show the allowed values of $\alpha$ as a function of $ (\sigma_{\gamma X} v_{\text{rel}})_{0}$
for a fixed $m_{\phi}=0.25\text{ GeV}$
with three different choices of cubic coupling  $\mu$. 
For sufficiently small $ (\sigma_{\gamma X} v_{\text{rel}})_{0}\ll 10^{-26}\mbox{ cm}^{3}\mbox{s}^{-1}$,
namely,  the total annihilation cross-section is dominated by
$(\sigma_{\phi\phi} v_{\text{rel}})_{0}$ at freeze out,
the  value of $\alpha$ is found to be insensitive to $ (\sigma_{\gamma X} v_{\text{rel}})_{0}$.
From the figure, one obtains $\alpha\approx 0.014$, $
0.016$, and $0.019$ for $\mu=0$, $m_\chi/2$, and $m_\chi$,
respectively.
When $(\sigma_{\gamma X} v_{\text{rel}})_{0}$ grows
and approaches $\sim 10^{-26}\mbox{ cm}^{3}\mbox{s}^{-1}$,
the value of $(\sigma_{\phi\phi} v_{\text{rel}})_{0} $ needs to be suppressed
in order to give the correct relic density,
which results in  a significant reduction of $\alpha$.
As it can be seen from the figure,
at $(\sigma_{\gamma X} v_{\text{rel}})_{0}\approx 10^{-26}\mbox{ cm}^{3}\mbox{s}^{-1}$, the
allowed values of $\alpha$ is reduced by $\sim 50\%$.

The effects of kinetic decoupling are shown in the right panel of Fig. \ref{fig:constraints}
for two different decoupling temperatures $T_{f}=2 T_{\text{kd}}$ and $10 T_{\text{kd}}$.
For small $ (\sigma_{\gamma X} v_{\text{rel}})_{0}\ll 10^{-26}\mbox{ cm}^{3}\mbox{s}^{-1}$,
the inclusion of kinetic decoupling only leads to a slight increase of $\alpha$,
which is expected as
in this region the $p$-wave annihilation dominates,
the kinetic decoupling makes DM particle freeze out more quickly,
which leads to a higher relic density
as discussed in the previous section.
For large $(\sigma_{\gamma X} v_{\text{rel}})_{0}\sim 10^{-26}\mbox{  cm}^{3}\mbox{s}^{-1}$,
the effect of kinetic decoupling is sizeable,
as in this region,
the $s$-wave annihilation becomes important,
which has a stronger dependence on the kinetic decoupling than that of $p$-wave case.
This leads to a significant decrease of $\alpha$.
Especially in the vicinity of resonances,
due to the additional $s$-wave resonant Sommerfeld enhancement,
there could be a sudden reduction on the size of  $\alpha$,
which can be seen  at
$(\sigma_{\gamma X} v_{\text{rel}})_{0}\sim 6\times 10^{-27}\mbox{  cm}^{3}\mbox{s}^{-1}$
in \Fig{fig:constraints}.

\begin{figure}
  \begin{center}
        \includegraphics[width=0.49\textwidth]{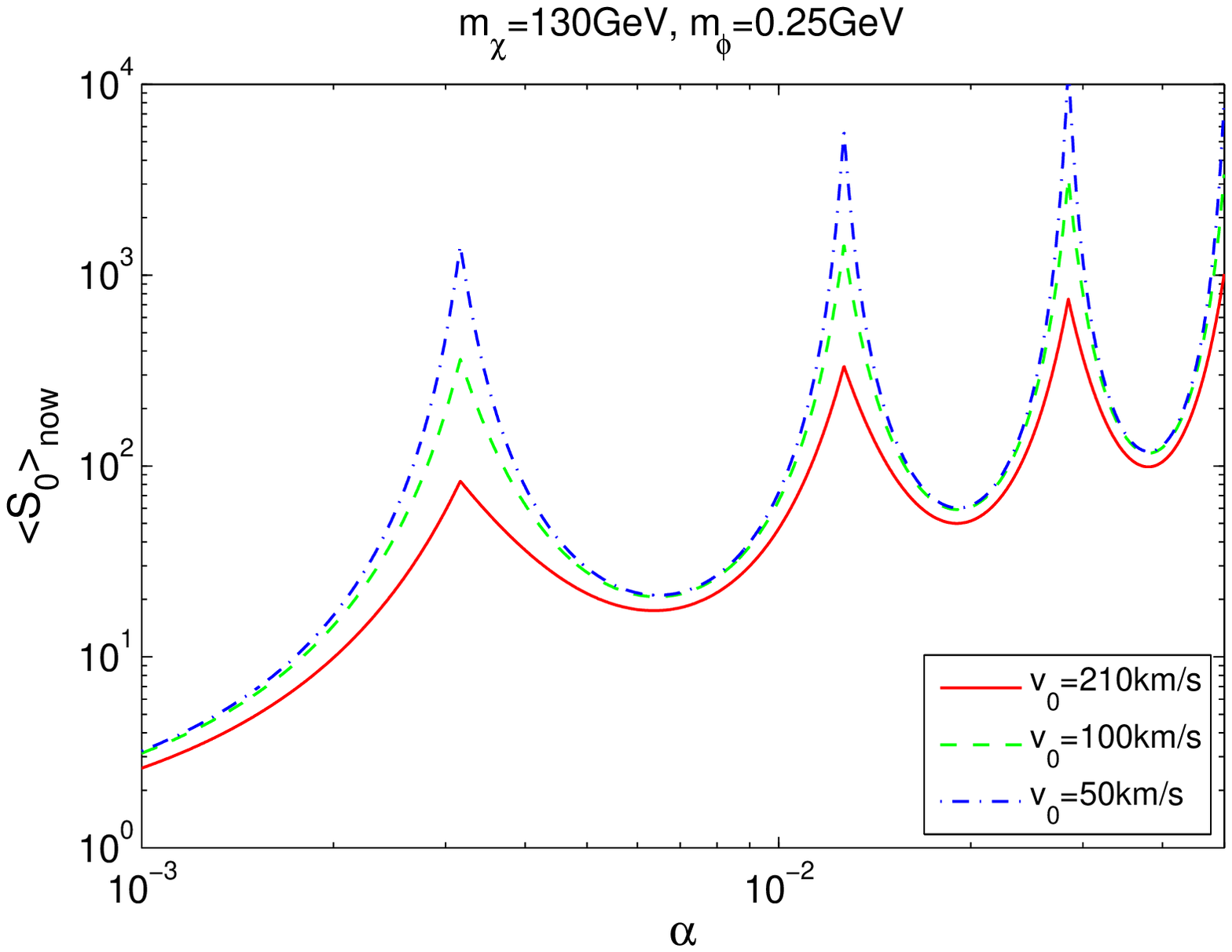}
          \includegraphics[width=0.49\textwidth]{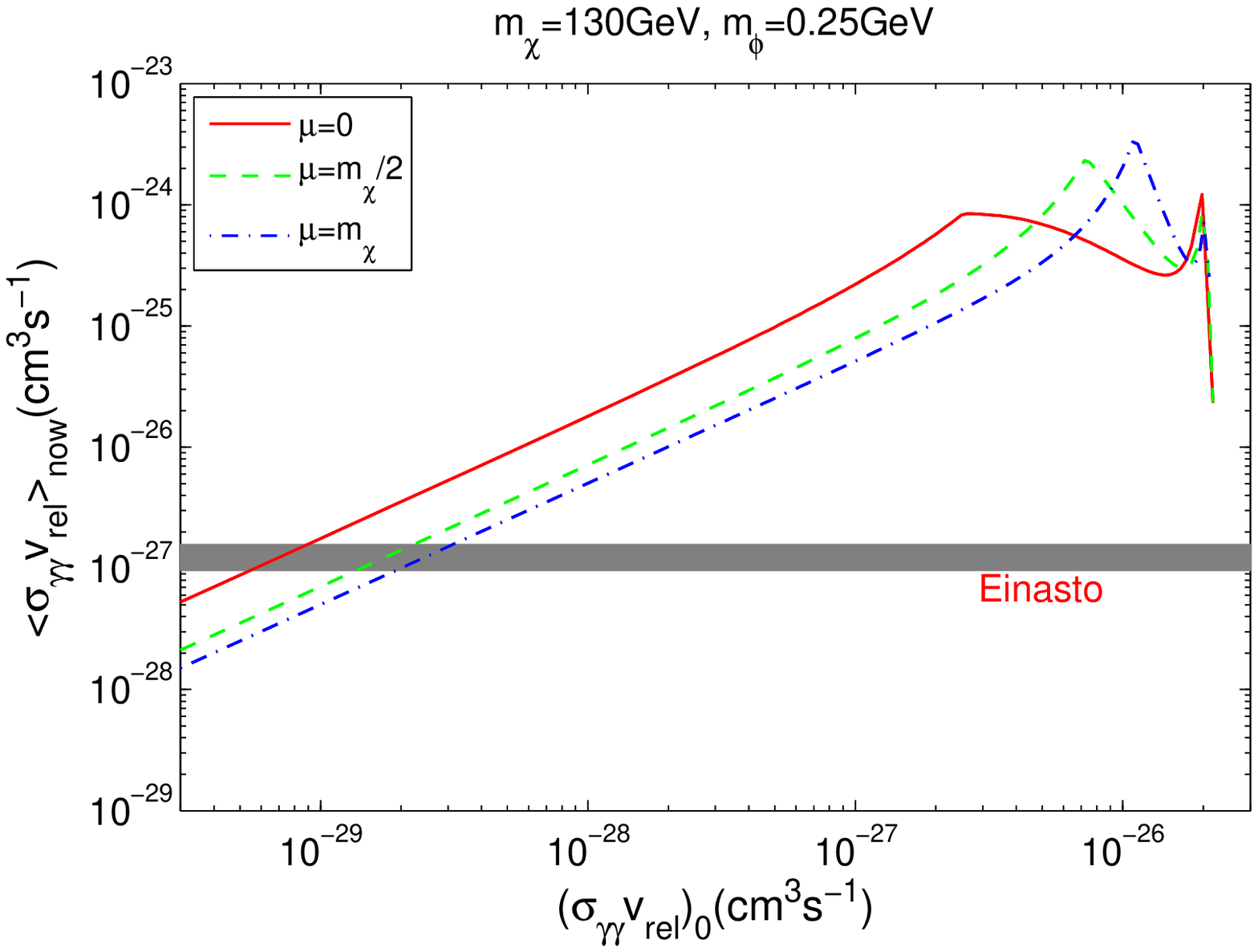}
          \caption{ (Left) $\langle S_{0}\rangle_{\text{now}}$ as a function
            of coupling strength $\alpha$ for the velocity dispersion
            $v_{0}=210\text{ km}\text{ s}^{-1}$, $100\text{ km}\text{ s}^{-1}$,
            and $ 50 \text{ km}\text{ s}^{-1}$ respectively. (Right) the
            relation between $\langle\sigma_{\gamma
              \gamma}v_{\text{rel}}\rangle_{\text{now}}$ and $(\sigma_{\gamma
              \gamma}v_{\text{rel}})_{0}$ for three different values of
            $\mu$. The gray band denotes the fit cross-section $\langle
            \sigma_{\gamma \gamma} v \rangle=(1.27\pm0.32)\times 10^{-27}\mbox{
              cm}^{3}\mbox{s}^{-1}$ in the case of Einasto
            profile~\cite{1204.2797:Weniger}. Other parameters are fixed at
            $m_{\chi}=130\text{ GeV}$ and $m_{\phi}=0.25\text{ GeV}$.  }
    \label{fig:constraints-smf}
\end{center}
\end{figure}

After constraining the allowed values of $\alpha$,
the $s$-wave Sommerfeld enhancements factor $\langle S_0\rangle$
at low temperatures can be calculated straightforwardly.
The velocity-averaged Sommerfeld enhancement of the halo DM annihilation cross-section
in the present-day is defined as
\begin{equation}
  \langle S_{0}\rangle_{\text{now}}=
  \frac{1}{N v_{0}^{3}}
  \sqrt{\frac{2}{\pi}}\int_{0}^{v_{\text{esc}}}S_{0} \
  e^{-\frac{v_{\text{rel}}^{2}}{2v_{0}^{2}}}v_{\text{rel}}^{2}dv_{\text{rel}} \ ,
\end{equation}
where $v_{0}$ is the DM velocity dispersion and $v_{esc}$ is the DM escape
velocity, respectively.
$N=\text{erf}\left(z/\sqrt{2}\right)-\left(2/\pi\right)^{1/2}ze^{-z^{2}/2}$ is a
normalization constant with $z=v_{\text{esc}}/v_{0}$.
Both $v_{esc}$ and $v_{0}$ depend on the distance $r$ from the GC.
In the vicinity of the Sun, $r=r_{\odot}\approx 8.5$ kpc,
$v_{\text{esc}}(r_{\odot})\approx 525\text{ km s}^{-1}$,
and $v_{0}(r_{\odot}) \approx 210\text{ km s}^{-1}$.
In the left
panel of Fig.~\ref{fig:constraints-smf},  we show the dependences of $ \langle
S_{0}\rangle_{\text{now}}$ on the coupling $\alpha$ and
the velocity dispersion $v_0$.
For the allowed values $\alpha=0.0136$, $ 0.0159$, and $0.0186$ corresponding to
$\mu=0$, $m_\chi/2$, and $m_\chi$, the enhancement factors are $\langle
S_{0}\rangle_{\text{now}}\approx 170$, $70$, and $50$, respectively.  Thus the
Sommerfeld enhancement factor can reach $\mathcal{O}(100)$ with the constraints from DM
thermal relic density, which is larger than the case where the force-carrier is a vector
boson~\cite{Feng:2009hw,Feng:2010zp}.
%

We are now in the position to
discuss the  line spectral shape  in the photon spectrum recently observed by
the Fermi-LAT collaboration.
The observation,
if interpreted as DM  annihilation into  $\gamma\gamma$,
corresponds to an $s$-wave velocity-averaged cross-section
$\langle\sigma_{\gamma\gamma} v_{\text{rel}}\rangle_{\text{now}}
\sim 10^{-27}\text{cm}^3\text{s}^{-1}$.
All of our previous  results on the annihilation $\chi\bar\chi \to \gamma X$
can be directly applied to the  case of  $\chi\bar\chi \to\gamma\gamma$,
as it is a special case of $\gamma X$.
In the right panel of \Fig{fig:constraints-smf},
the relation between $\langle \sigma_{\gamma \gamma} v_\text{rel}\rangle_\text{now}$ and
$(\sigma_{\gamma \gamma} v_\text{rel})_0$ is shown.
One sees that in order to reproduce
the observed signal corresponding to
$\langle \sigma_{\gamma\gamma} v_{\text{rel}}\rangle_{\text{now}} \approx 1.27\times 10^{-27}\text{cm}^3\text{s}^{-1}$
for the Einasto profile~\cite{1204.2797:Weniger},
the required cross-sections before the
Sommerfeld enhancement can be quite  small
$(\sigma_{\gamma\gamma} v_{\text{rel}})_0$=$7.2\times 10^{-30}\mbox{ cm}^{3}\mbox{s}^{-1}$,
$1.8\times 10^{-29}\mbox{ cm}^{3}\mbox{s}^{-1}$,
and $2.5\times 10^{-29}\mbox{ cm}^{3}\mbox{s}^{-1}$,
for $\mu=0$, $m_\chi/2$, and $m_\chi$, respectively.

The cross-section of $\chi\bar\chi \to \gamma\gamma$ may be further enhanced
due to the possible lower DM velocity dispersion near the GC than
that in the solar neighborhood.
The dependence of Sommerfeld enhancement on $v_{0}(r)$  at GC has been discussed in
 Refs.~\cite{Cirelli:2010nh,Abazajian:2011ak}.
 The N-body simulations suggest that $v_{0}(r)$ is related to the DM density profile $\rho(r)$ through
 a relation
 $v_{0}(r)^{3}/\rho(r)\approx r^{\chi} $~\cite{Bertschinger1985}.
 From pure DM simulations,
 the power index is found to be $\chi\simeq 1.9-2.0$~\cite{0810.1522:Navarro}.
 For the NFW profile
 $\rho(r)\propto (r/r_{s})^{-\alpha}(1+r/r_{s})^{-3+\alpha}$
 with $\alpha=1.0$ and $r_{s}=20$ kpc,
 $v_{0}(r)$ scales as $r^{\chi-\alpha}$.
 Therefore it decreases with the distance $r$.
 This is also true for the Einasto profile
 $\rho(r)\propto\exp[-(2/\alpha)(r/r_{s})^{\alpha}]$,
 if one takes the typical value of $\alpha=0.17$.
 Thus a larger $s$-wave Sommerfeld enhancement at GC can be expected.
 However, if the Sommerfeld enhancement saturates
 at some velocity smaller than  $v_{0}(r_\odot)$,
 then it is insensitive to the choice of $v_{0}(r)$.
 The additional enhancement at GC in this case can be simply  estimated as
 $S_{0}(v=0)/S_{0}(v_{0}(r_{\odot}))$~\cite{1107.3546}.
Note that the simulation results can be modified significantly after
including baryons \cite{0911.2316_White}.
%

\begin{figure}[htb]
   \begin{center}
     \includegraphics[width=0.6\textwidth]{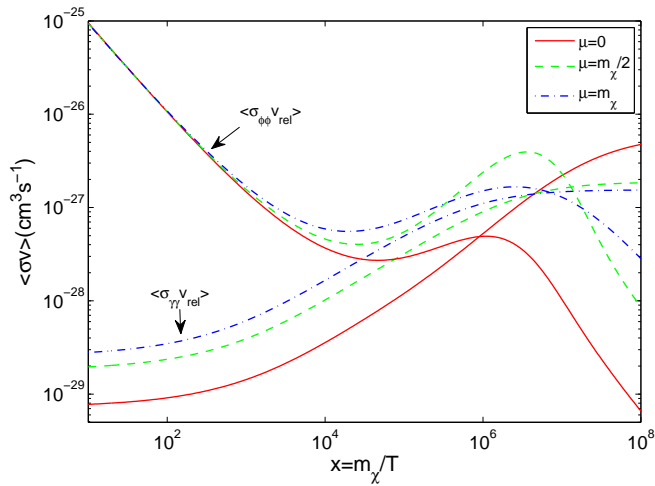}
     \caption{Temperature evolution of $\langle \sigma_{\phi\phi}
       v_{\text{rel}}\rangle$ and $\langle \sigma_{\gamma\gamma}
       v_{\text{rel}}\rangle$ for three different values of $\mu=0$,
       $m_\chi/2$, and $m_\chi$, respectively. }
     \label{fig:thermal-average}
\end{center}
\end{figure}

Although at freeze out the $p$-wave cross-section
$\langle \sigma_{\phi\phi} v_{\text{rel}}\rangle$ can be a few order of magnitudes larger than
the $s$-wave cross-section $\langle \sigma_{\gamma\gamma}v_{\text{rel}}\rangle$,
at lower temperatures it is possible that
$\langle \sigma_{\phi\phi} v_{\text{rel}}\rangle$
becomes  comparable with or even smaller  than
$\langle\sigma_{\gamma\gamma} v_{\text{rel}}\rangle$,
which is due to
the dramatic difference in velocity-dependencies between
$s$- and $p$-wave processes in the presence of Sommerfeld enhancement.
This possibility is shown  in \Fig{fig:thermal-average}.
%
At the temperature of thermal decoupling $x\approx25$,
one sees that
$\langle\sigma_{\phi\phi} v_{\text{rel}}\rangle\approx 1\times10^{-25}\text{ cm}^{3}\text{s}^{-1}$
 which is about four order of magnitudes larger than
$\langle \sigma_{\gamma\gamma} v_{\text{rel}}\rangle$.
When the temperature goes down,
in general
the value of $\langle \sigma_{\phi\phi} v_{\text{rel}}\rangle$ decreases due to the
velocity-suppression.
The $p$-wave Sommerfeld enhancement of $\langle \sigma_{\phi\phi}
v_{\text{rel}}\rangle$ can be significant in the range $10^{4}\lesssim x \lesssim 10^{6}$
when it is near a resonance region.
But for other values of $x$ which are off the resonance,
$\langle \sigma_{\phi\phi} v_{\text{rel}}\rangle$ decrease rapidly.
On the other hand,
the size of $\langle \sigma_{\gamma\gamma} v_{\text{rel}}\rangle$
increases monotonically towards larger $x$ due to
the $s$-wave Sommerfeld enhancement.
At $x\approx 25$,
the value of $\langle \sigma_{\gamma\gamma} v_{\text{rel}}\rangle$ is
$\sim 10^{-29}\text{ cm}^{3}\text{s}^{-1}$,
but at $x\approx4\times 10^6$ which corresponds to $v_0=210\text{ km}\text{ s}^{-1}$,
it reaches $\sim 10^{-27}\text{ cm}^{3}\text{s}^{-1}$ and becomes comparable with
$\langle\sigma_{\phi\phi} v_{\text{rel}}\rangle$ for $\mu=m_\chi/2$ and $m_\chi$.
For $\mu=0$ case, it can even dominate over $\langle\sigma_{\phi\phi} v_{\text{rel}}\rangle$.
Therefore,
it is  possible that $\chi\bar\chi\to \gamma\gamma$ can be the
main DM annihilation channel today,
and is responsible for the observed gamma-ray line at the GC.
%
%
In this work, we consider a light scalar with $m_\phi\approx 0.25$ GeV,
such that $\phi$ can decay into $\mu^+\mu^-$ and $e^+ e^-$.
But the final states of $\pi^0\pi^0$ and $\tau^+\tau^-$ are kinematically forbidden,
which will suppress the generation of  large continuum gamma-ray flux from the annihilation
$\chi\bar\chi\to \phi\phi$  followed by the decay of $\phi$ into these final states.

Note that such a large $s$-wave enhancement of $\mathcal{O}(100)$ is still consistent with
the limits derived from BBN and CMB data \cite{1102.4658},
as the total DM annihilation cross-section after the Sommerfeld enhancement
is still an order of magnitude smaller than the typical WIMP annihilation cross-section.

\section{Sommerfeld enhancement  in a  simple model}\label{sec:model}

We have shown in the previous section that
the  Sommerfeld-enhanced cross-section
$\langle\sigma_{\gamma\gamma} v_{\text{rel}}\rangle \sim 10^{-27}\text{ cm}^{3}\text{s}^{-1}$
today can be consistent with
the DM thermal relic density, due to the $p$-wave annihilation of DM particles into the force-carriers.
In this section,
we discuss another advantage of  the Sommerfeld enhancement,
namely, it  can also solve the problem of
unnaturally large couplings required  by many DM models
motivated to explain the putative gamma-ray line.
As the DM particles carry  no electric charge,
it is often assumed that
the annihilation of  DM particles  into $\gamma\gamma$ proceeds only through
one-loop diagrams which involve
massive charged particles running in the loop
( for exceptions, see e.g.  Refs. \cite{1205.1520_Mambrini,1209.1093_Weiner} ).
Due to the loop suppression, in general
the effective  DM couplings to the charged particles  in the loop
have to be above order unity,
in order to give  the cross-section suggested by Fermi-LAT (see e.g. \cite{1205.6811:Hooper}),
which may raise the issue of perturbativity.
Invoking  the mechanism of Sommerfeld enhancements
can significantly reduce
the required couplings to the  perturbative region.

\begin{figure}
  \begin{center}
    \includegraphics[width=0.5\textwidth]{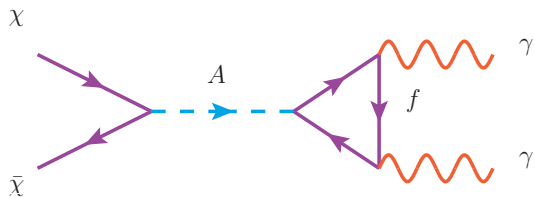}
    \caption{Feynman diagram for $\chi\bar{\chi}\to\gamma\gamma$ through loop diagrams with an intermediate pseudoscalar $A^0$ and a charged fermion $f$ in the loop.}\label{fig:one-loop}
\end{center}
\end{figure}

For a concrete illustration,
we consider a realization of Sommerfeld enhancement
in a  simple reference model.
In this  model,
the DM particle $\chi$ is assumed to be a Majorana fermion.
Other particles in the models are:
 a light scalar force-carrier $\phi$,
 a heavy pseudoscalar mediator $A^{0}$,
 and a heavy charged Dirac fermion $f$ with
 electromagnetic charge number  $Q_{f}$ and color number $C_{f}$.
 The relevant interactions in the model are given by
 the following Lagrangian
\begin{equation}
\mathcal{L}_{int} \subset -\frac{g}{2}\bar{\chi}\phi\chi-i \frac{g_{\chi}}{2} \bar{\chi}\gamma_{5}\chi A^{0}-ig_{f}\bar{f}\gamma_{5}f A^{0} .
\end{equation}
In order to avoid constraints from the nonobservation of continuum photon spectrum,
we consider the case where  $m_{f}, m_{A} \gg m_{\chi}$,
such that $\chi\bar\chi\to f \bar f, A^0A^0$ are kinematically forbidden,
and
$\chi\bar{\chi}$ can only annihilate into $\gamma\gamma$ through loop process.
The corresponding Feynman diagram is  shown in  Fig.~\ref{fig:one-loop}.
The annihilation cross-section for this process is given by \cite{1208.0009:Zurek}
\begin{equation} \label{eq:csto2gamma}
\left(\sigma_{\gamma\gamma} v_{\text{rel}}\right)_0=\frac{1}{4\pi^{3}}\frac{\alpha_{\text{em}}^{2}g_{\chi}^{2}g_{f}^{2}Q_{f}^{4}C_{f}^{2}m_{f}^{2}}{(s-m_{A}^{2})^{2}+m_{A}^{2}\Gamma_{A}^{2}}\left[\arctan\left(\frac{1}{\sqrt{m_{f}^{2}/m_{\chi} ^{2}-1}}\right)\right]^{4} ,
\end{equation}
where $s$ is the center-of-mass total energy.
The total width $\Gamma_A$ of
the pseudoscalar $A^0$ receives contribution from the decay channels
$A^0\to \chi\bar\chi$, $f\bar f$, and $\gamma\gamma$ with
the  partial widths
 \begin{eqnarray}\label{eq:width}
\Gamma_{\chi \bar \chi}=
\frac{g_{\chi}^2 m_A}{16\pi} \sqrt{1-\frac{4m_{\chi}^2}{m_A^2}} ,
 \ \Gamma_{f\bar f}=
\frac{g_{f}^2 m_A}{8\pi} \sqrt{1-\frac{4m_{f}^2}{m_A^2}} ,
\end{eqnarray}
and
\begin{equation}
 \Gamma_{\gamma\gamma}=\frac{m_{A}^{3}\alpha_{em}^{2}g_{f}^{2}Q_{f}^{4}C_{f}^{2}}{256\pi^{3}m_{f}^{2}}
  |A^{A}_{1/2}(m_{A}^{2}/4m_{f}^{2})|^{2} ,
\end{equation}
where the function $A^{A}_{1/2}$ is defined as $A^{A}_{1/2}(\tau)=2\tau^{-1}f(\tau)$ with
\begin{equation}
  f(\tau)=\left\{
    \begin{array}{ll}
      \arcsin^2\sqrt{\tau}  & (\text{for } \tau \leq 1) \nonumber\\
      -\frac{1}{4}\left(\ln \frac{1+\sqrt{1-1/\tau}}{1-\sqrt{1-1/\tau}}-i\pi \right)^{2} & (\text{for } \tau >1)
    \end{array}
    \right.  .
  \end{equation}
We assume  that
in the dark  sector the parity symmetry is conserved such that
the decay $A^{0} \to \phi\phi$ is forbidden.
Since $A^{0}$ is a pseudoscalar,
the annihilation $\chi\bar\chi\to \gamma\gamma$ proceeds through $s$-wave.
For heavy $m_A$, $m_f\gg m_\chi$,
the size of the annihilation cross-section can be
estimated as
\begin{equation}
\left(\sigma_{\gamma\gamma} v_{\text{rel}}\right)_0 \sim
10^{-27} \text{cm}^{3}\text{s}^{-1} \left( \frac{g_\chi g_f}{100}\right)^2 \left( \frac{m_\chi}{130\mbox{ GeV}}\right)^4
\left( \frac{ 500 \mbox{ GeV} }{ m_A}\right)^4\left( \frac{ 500 \mbox{ GeV} }{ m_f}\right)^2 .
\end{equation}
One sees that
a very large product of the couplings $\sqrt{g_\chi g_f}\sim 10$ is required,
which is unnatural and can invalidate  the perturbative calculations.
This is a common problem in many dark matter models in which the DM annihilation into
$\gamma\gamma$ through one-loop diagrams.

One way to enhance the cross-section  is to assume that the annihilation is near a resonance which occurs if
$m_A\approx 2 m_\chi+\delta$ with $\delta\ll m_A$. In this case the cross-section can be
 enhanced by a  factor of $m_A^2/(4 \delta^2 + \Gamma_A^2)$.
 If small couplings
 $g_{\chi}, g_f\approx1$ is required, the relative mass difference should be around $\delta/m_A\approx 2\%$ for $\Gamma_A\approx \Gamma_{\chi\bar\chi}$.
 Another modest enhancement may arise from the case where $m_\chi\approx m_f$ such that
the  function $\arctan[(m_f^2/m_\chi^2-1)^{-1/2}]$ reaches its maximal value $\pi/2$. Compared with the
case where $m_f\approx 5 m_\chi$ the enhancement of the value of the arctan is $\sim 8$. In addition, in order to
have the correct relic density, it is required that $\delta$ must be positive \cite{1208.4100:Bai}.

\begin{figure}
  \begin{center}
    \includegraphics[width=0.65\textwidth]{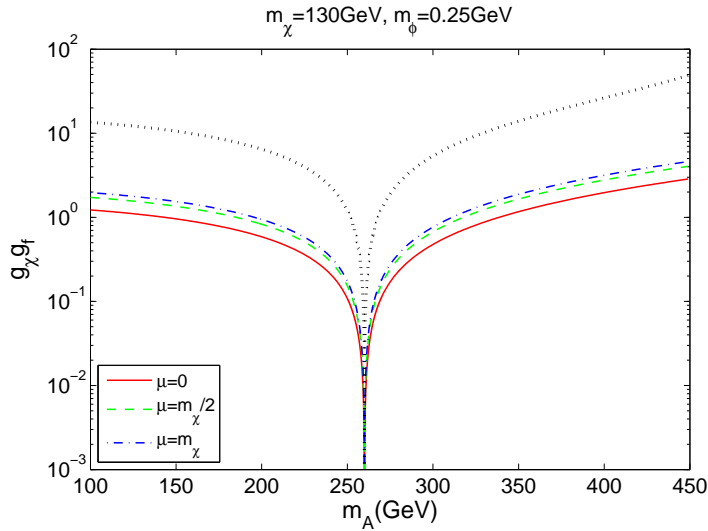}
    \caption{
      The required value of $g_{\chi}g_{f}$ to account for the gamma-ray line signals with cross-section
      $\langle\sigma_{\gamma \gamma} v_{\text{rel}} \rangle=1.27\times 10^{-27}\mbox{cm}^{3}\mbox{s}^{-1}$~\cite{1204.2797:Weniger}.
      Three cases  with Sommerfeld enhancement  correspond to $\mu=0$ (solid), $m_\chi/2$ (dashed), and $m_\chi$ (dot-dashed),
      respectively.   The dotted line denotes the case without Sommerfeld enhancement.  Other parameters are
      fixed at $Q_{f}=C_{f}=1$, $g_{\chi}=g_{f}$, and $ m_{f}=300\GeV$.
    }
    \label{fig:coupling}
\end{center}
\end{figure}

With the presence  of Sommerfeld enhancements, the required couplings can be reduced
significantly without introducing any mass degeneracies.
In Fig. \ref{fig:coupling}, we show the required couplings which can reproduce
the observed
$\langle \sigma_{\gamma\gamma} v_{\text{rel}}\rangle=1.27\times 10^{-27} \text{ cm}^3\text{s}^{-1}$. Compared with the case without Sommerfeld enhancements, the required products
$g_\chi g_f$ can be reduced by an order of magnitude. For a wide range of the pseudoscalar mass
$100<m_{A}<450$ GeV, the required couplings are smaller than unity. If the Sommerfeld enhancements are absent, only a narrow region close to $2m_\chi$ can be consistent with a small
$g_\chi g_f$.

In this simple model,
the force-carrier $\phi$ does not appear  in
the leading one-loop diagram for
$\chi\bar\chi\to\gamma\gamma$,
which make it straightforward to
factorize the short- and long-distance contributions.
In general,
the force-carrier can play  important roles  in
both the one-loop diagram  and the long-distance Sommerfeld enhancements.
For instance,
in the case where
the DM particle is a Majorana fermion
belonging to the adjoint presentation of $SU(2)_W$ gauge group
(e.g  Wino neutralino),
the $SU(2)_W$ gauge bosons $W^\pm$ appear
in both the  box-diagrams for DM annihilation into $\gamma\gamma$
and the long-distance ladder diagrams as a force-carrier which
results in off-diagonal Sommerfeld enhancements
\cite{Hisano:2002fk,Hisano:2003ec,Hisano:2004ds,Cirelli:2007xd,Hryczuk:2010zi,Hryczuk:2011tq}.
A  fully consistent analysis of incorporating these two contribution simultaneously can be found
in Ref.\cite{Hryczuk:2011vi}.


\section{Conclusions}\label{sec:conclusion}
In summary, the recently reported possible indications of line spectral
features in the Fermi-LAT photon spectrum in regions close to the galactic
center can be related to the signals of halo DM annihilation. However, the
corresponding annihilation cross-section of $\langle\sigma_{\gamma\gamma}
v_{\text{rel}}\rangle\approx\mathcal{O}(10^{-27})\text{ cm}^{3}\text{s}^{-1}$
is too large for typical loop-induced radiative processes, while on the other
hand too small to be responsible for the observed DM relic density which is
typically $3\times10^{-26}\text{ cm}^{3}\text{s}^{-1}$.  We have pointed out
that the Sommerfeld enhancement with scalar force-carrier can simultaneously
explain those features.  In this mechanism, the cross-section for DM
annihilation into $\gamma X$ in the galactic center today can be greatly
enhanced due to the attractive forces between the DM particles, which is
induced by the multiple-exchange of the force-carriers. The additional
$p$-wave annihilation into the force-carriers can dominate the total DM
annihilation cross-section at freeze out and set the correct thermal relic
density, but can have subdominant contributions to the DM annihilation today
due to the velocity suppression.  We have performed an detailed analysis on
the allowed Sommerfeld enhancement under the constraints from the thermal
relic density determined by $p$-wave annihilation processes.  The results show
that the Sommerfeld enhancement factor can reach $\mathcal{O}(100)$.  In a
simple reference model with $\chi\bar\chi\to 2\gamma$ occurring at one loop, we
have shown that the required DM effective couplings to the intermediate
particles in the loop can be reduced by an order of magnitude and below
unity, which keeps the perturbativity of the model.  Compared with some other mechanisms
for increasing the DM annihilation cross-section at loop level, the Sommerfeld
enhancement does not require any degeneracies in the masses of DM particles
and the intermediate particles in the loop diagrams.

\subsection*{Acknowledgments}
This work is supported in part by
the National Basic Research
Program of China (973 Program) under Grants No. 2010CB833000;
the National Nature Science Foundation of China (NSFC) under Grants
No. 10975170,
No. 10821504 and No. 10905084;
and the Project of Knowledge Innovation Program
(PKIP) of the Chinese Academy of Science.


\providecommand{\href}[2]{#2}\begingroup\raggedright\endgroup

\end{document}